\documentclass[graybox]{svmult}

\usepackage{type1cm}        % activate if the above 3 fonts are
\usepackage{makeidx}         % allows index generation
\usepackage{graphicx}        % standard LaTeX graphics tool
\usepackage{subcaption}                             % when including figure files
\usepackage{multicol}        % used for the two-column index
\usepackage[bottom]{footmisc}% places footnotes at page bottom
\usepackage{booktabs}

\usepackage[numbers]{natbib}
\usepackage{newtxtext}       % 
\usepackage{newtxmath}       % selects Times Roman as basic font

\usepackage{mathtools}

\DeclareMathOperator*{\argmin}{arg\,min}

\graphicspath{ {./images/} }
\newcommand{\F}{\mathcal{F}}
\newcommand{\Pk}{\mathcal{P}}

\newcommand{\T}{\mathcal{T}}
\usepackage{dsfont}
\usepackage{cleveref}
\newcommand{\essp}{\operatorname{ess}\operatorname{sup}}
\newcommand{\essf}{\operatorname{ess}\operatorname{inf}}
\usepackage{pifont}
\newcommand{\cmark}{{\color{green}\ding{51}}}%
\newcommand{\xmark}{{\color{red}\ding{55}}}%

\makeindex             % used for the subject index
\begin{document}

\title*{Decision-Dominant Strategic Defense Against Lateral Movement for  5G Zero-Trust Multi-Domain Networks}
\titlerunning{Decision-Dominant Zeto-Trust Defense in Multi-Domain Networks}
\author{Tao Li, Yunian Pan, and Quanyan Zhu}
\institute{Tao Li (Corresponding author) \at New York University, NY, 11201, \email{tl2636@nyu.edu}
\and Yunian Pan \at New York University, NY, 11201 \email{yp1170@nyu.edu}
\and Quanyan Zhu \at New York University, NY, 11201 \email{qz494@nyu.edu}
}
\maketitle

\abstract{Multi-domain warfare is a military doctrine that leverages capabilities from different domains, including air, land, sea, space, and cyberspace, to create a highly interconnected battle network that is difficult for adversaries to disrupt or defeat. However, the adoption of 5G technologies in battlefields presents new vulnerabilities due to the complexity of interconnections and the diversity of software, hardware, and devices from different supply chains. Therefore, establishing a zero-trust architecture for 5G-enabled networks is crucial for continuous monitoring and fast data analytics to protect against targeted attacks. To address these challenges, we propose a proactive end-to-end security scheme that utilizes a 5G satellite-guided air-ground network. Our approach incorporates a decision-dominant learning-based method that can thwart the lateral movement of adversaries targeting critical assets on the battlefield before they can conduct reconnaissance or gain necessary access or credentials. We demonstrate the effectiveness of our game-theoretic design, which uses a meta-learning framework to enable zero-trust monitoring and decision-dominant defense against attackers in emerging multi-domain battlefield networks.}

{
\section{Introduction}
\label{sec:intro}
The U.S. military has been undergoing a doctrine transition from traditional single to multi-domain operations or warfare (MDW), which the Army formally approved in October 2022 as its new warfighting doctrine \cite{FM}. The new doctrine defines MDW as ``the combined arms employment of joint and Army capabilities to create and exploit relative advantages that achieve objectives, defeat enemy forces, and consolidate gains on behalf of joint force commanders,'' \cite{FM} which directs the service to combine and integrate air, land, sea, space, and cyberspace in all facets of operations. MDW is developed in response to the 2018 National Defense Strategy \cite{2018defense}, shifting the previous focus of U.S. national security from addressing violent extremists worldwide to great power competition and potential conflict with near-peer adversaries across air, land, sea, space, and cyberspace. 

One main impetus for this doctrine transition is the technological advances and increased complexity of modern warfare. In addition to traditional platforms such as main battle tanks and guided-missile destroyers, the rise of space, information, and artificial intelligence technologies leads to enhanced and new military capabilities, such as the Advanced Extremely High-Frequency Systems \cite{aehf} powered by military satellites, the Indago quadrotor unmanned aerial systems \cite{indagouav}, and the U.S. cyber force. By leveraging the strengths of various military capabilities across multiple domains, military forces operate through the physical dimension (air, land, sea, space), influence through the information dimension (cyberspace), and achieve victory in the human dimension.   

MDW involves seamless coordination and integration of forces and assets across domains to gain a competitive advantage over adversaries. For example, ground forces may work in conjunction with air and space assets to gain situational awareness, conduct precision strikes, and provide close air support. Meanwhile, naval forces may coordinate with cyberspace capabilities to disrupt an adversary's communication networks and gain information superiority. The fifth-generation (5G) wireless technology plays an important role in MDW because it provides a network infrastructure that enables faster data transfer, greater bandwidth, lower latency, and increased capacity compared to its predecessors. With 5G networks, military units across multiple domains can access and share information in real time, creating a synergistic effect that improves situational awareness and enhances command and control. Furthermore, 5G connectivity can facilitate the communication and control of unmanned and autonomous systems powered by artificial intelligence both on the ground and in the air, enabling the integration of unmanned assets into MDW. A schematic illustration of 5G networks in MDW is presented in Figure~\ref{fig:5g} 
\begin{figure}
    \centering
    \includegraphics[width=0.9\textwidth]{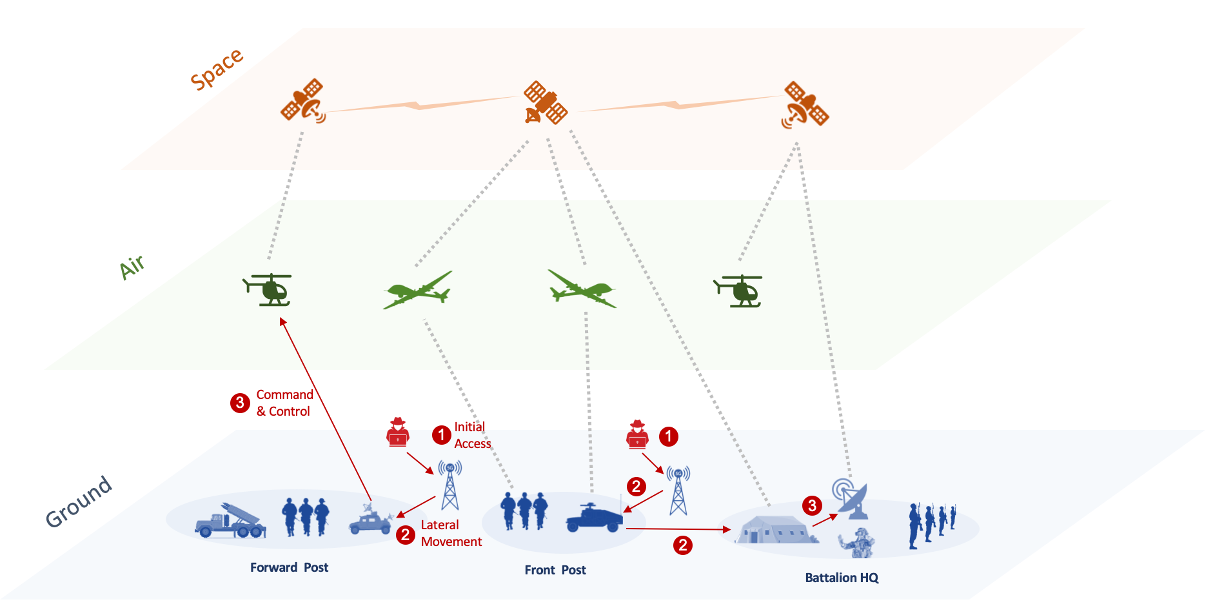}
    \caption{An illustration of 5G Multi-Domain Networks (MDN). The army force has deployed a robust 5G communication infrastructure to facilitate seamless communication within the base and between the battalion headquarters (HQ), front lines, and forward posts. Additionally, the integration of 5G-powered satellites enables effective communication between aerial vehicles and ground forces. An APT attack can start with initial access (1), create lateral movement (2), and eventually command and control the targeted assets (3). Several paths of the attack chain are depicted, leading to the consequence of the compromise of a helicopter or misdirection to satellites}
    \label{fig:5g}
\end{figure}

Recent years have seen the adoption and implementation of 5G networks for military applications gaining momentum. The advanced features of 5G networks, despite their contributions to coordinated MDW operations, introduce security challenges periling the efficiency and effectiveness of MDW. For example, with more devices and sensors connected to the network system, 5G networks present a larger attack surface, e.g., more potential entry points for attackers to exploit, compared to previous generations. Meanwhile, as 5G networks provide faster and more reliable connectivity, they enable more sophisticated cyberattacks, such as large-scale distributed denial-of-service attacks \cite{huang2022radams}, network slicing exploitation \cite{slicing}, and edge computing compromise \cite{edge}.       

Among these cyberattacks, one critical threat is the Advanced Persistent Threat (APT).  APT attacks are typically carried out by skilled and well-funded attackers who use sophisticated techniques to gain unauthorized access to sensitive information and systems. APT attackers may conduct extensive network reconnaissance to gather information about the 5G network and its vulnerabilities. They exploit vulnerabilities in the 5G network and gain unauthorized access to a device or system within the network to move laterally through the network and access other devices or systems within it. In 5G networks, lateral movement capabilities can be particularly dangerous, as they can allow attackers to gain access to critical systems and data within the network. For example, an attacker who gains access to a single device within a 5G network could potentially use lateral movement techniques to access other devices or systems, such as servers or databases containing sensitive or confidential data.

Since military assets and systems across various domains are connected and rely on 5G networks to exchange information and coordinate operations, the vulnerability of 5G networks can pose significant challenges in MDW. Therefore, military organizations shall prioritize the security of 5G networks in MDW and establish a proactive cyber defense in 5G networks. The primary objective of such a cyber defense is to disrupt the attacker's kill chain, which includes the following stages: reconnaissance, privilege escalation, exploitation, lateral movement, and command and control. Starting from an entry point, the attacker gains initial access to the network, conducts reconnaissance, stealthily navigates within the 5G infrastructure, and ultimately compromises the targeted asset, such as a drone or a satellite. Such adversarial behaviors are increasingly common in APTs. %These adversaries, often backed by nation states, demonstrate high resourcefulness. They operate stealthily, actively seeking opportunities to gain unauthorized access, move within networks, and aim to control targeted assets. 5G networks feature numerous API interfaces and a diverse supply chain. They will have many unknown vulnerabilities, which can be exploited by APT adversaries.

To counteract the attacker's actions, the defender employs a sequence of defense actions known as the cyber defense chain, including monitoring, detection, response, and attribution. Figure~\ref{fig:kill-chain} summarizes the kill and the defense chains. The relationship between the kill and the defense chains is competitive in nature. The kill chain aims to evade the detection from the cyber chain to reach the target, while the defense chain aims to thwart the attack before an adversary carries out the planned attack. To outmaneuver the adversary's decision-making cycle, a defender needs superior situational awareness together with fast and reliable reasoning capabilities, especially in unknown and uncertain situations to make timely and effective decisions. These desiderata are also known as decision dominance. Illustrated in Figure~\ref{fig:kill-chain}, a decision-dominant defense at the monitoring and detection stage has the capability of gathering, processing, and analyzing information from various sources to obtain a comprehensive understanding of the cyber operational environment. At the response stage, a decision-dominant defense can quickly evaluate available options, assess risks, and make informed decisions in a timely manner. As a result, it thwarts the planned attack before its execution. To achieve decision dominance, there is a need for proactive cyber mechanisms, such as cyber deception and attack engagement, to gather immediate intelligence. In addition, agility is indispensable. It allows the defender to learn, adapt, and respond to changing situations, seize opportunities, and effectively adjust strategies and tactics as required. Strategic thinking is paramount to achieving agility, involving the study of adversarial behaviors, the development of adaptive tactics, and the ability to make informed and decisive decisions.

% and an adversarial attack is thwarted when the defender can promptly and accurately detect and respond to the attacker.
\begin{figure}[!ht]
    \centering
    \includegraphics[width=0.8\textwidth]{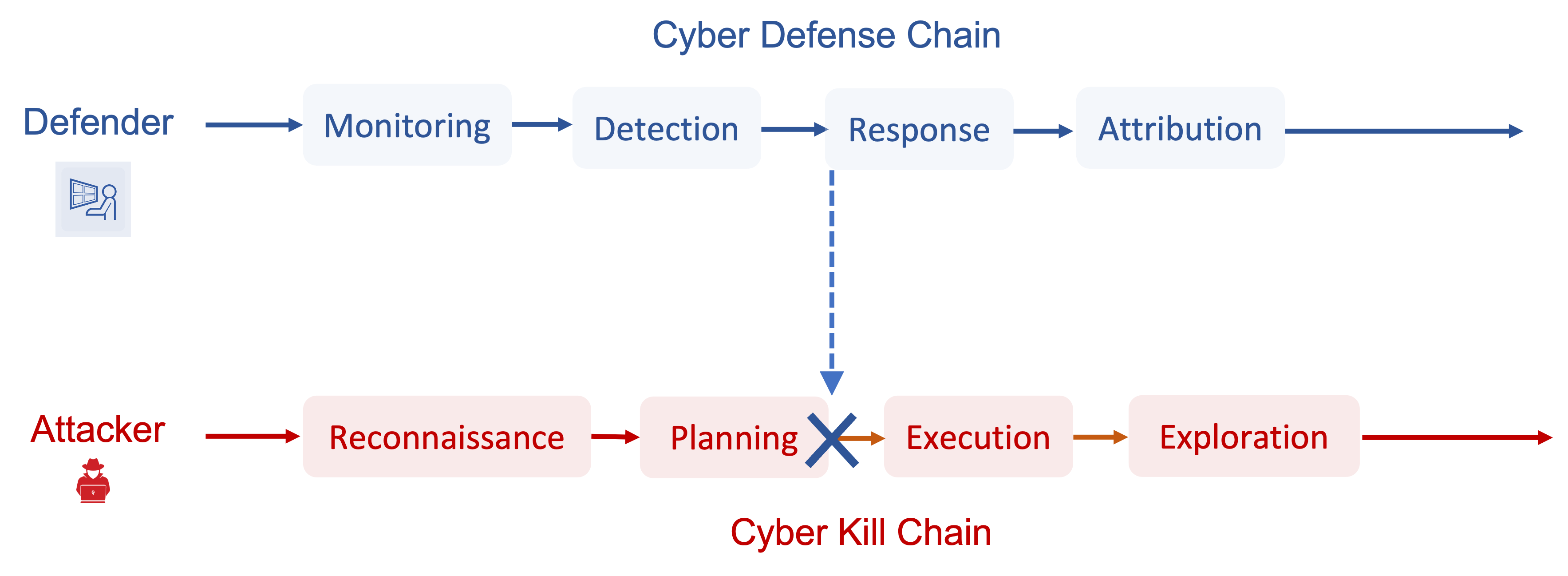}
    \caption{An illustration of cyber kill/defense chains. The kill chain consists of crucial stages such as reconnaissance, planning, execution, and exploration. The objective of defense measures is to disrupt this kill chain by employing monitoring, detection, response, and attribution techniques. An effective defense strategy is considered decision-dominant when it efficiently acquires and processes information, enabling it to make timely decisions that outpace the attacker. For instance, the defense chain can respond swiftly to thwart the attack even before the attacker initiates the planned offensive actions}
    \label{fig:kill-chain}
\end{figure}

There is a pressing need for the development of a systematic approach to establishing decision-dominant mechanisms for the defense of 5G networks. Game theory offers a promising solution in this regard. Not only does game theory naturally provide a framework for designing tactics in competitive environments, but recent advancements in dynamic games, learning, and their intersection with modern machine learning techniques enhance the reasoning capabilities of agents. This enables a formal and agile approach to achieving rapid decision-making.
For instance, {recent studies} \cite{zhu2018multi,huang2020dynamic,rass2016gadapt,huang2019dynamic} have introduced a class of dynamic games that effectively capture the evolving interactions between defense and kill chains. The concept of non-equilibrium has been proposed to derive solution concepts based on players' behaviors. This concept holds significant implications for cybersecurity applications, particularly when the interactions between attackers and defenders may be limited and indirect.

Another significant advantage of utilizing game-theoretic models is their strong epistemic foundation, which allows for explicit modeling and analysis of scenarios involving information asymmetry and the pace of decision-making. These models find wide applicability in 5G security networks. Information asymmetry arises from the fact that neither party possesses a comprehensive view of the entire 5G network. Instead, each party gathers partial observations through reconnaissance (the attacker) or monitoring (the defender). To effectively outmaneuver the adversary, the defender must establish an information advantage by actively acquiring information during the monitoring process. This proactive approach enables the defender to gain high-confidence situational awareness of the network system and adversarial behaviors. However, it is important to note that having an information advantage alone does not necessarily guarantee the defender an upper hand in cyber defense. Another crucial aspect that holds equal importance is the pace of decision-making. The defender faces a disadvantage if the attacker manages to execute the attack successfully before an adequate response can be mounted. In this regard, game theory frameworks provide a means to comprehensively capture the end-to-end decision-making process, encompassing information acquisition, learning, and decision-making. It provides a theoretical underpinning for understanding the fundamental tradeoff among these factors and a holistic approach to modeling and devising tactics across all stages.

One implicit assumption underlying the defense against APTs is that the attacker possesses the necessary capabilities to acquire initial access and credentials, and then establish a foothold within the network. We cannot stop the attack from getting into the network. This assumption forms the basis of the zero-trust security doctrine, which emphasizes the need to trust no entity by default and requires organizations to verify and authenticate all users, devices, and activities, regardless of their location or origin.
Recognizing the importance of assuming a reasonable capability of adversaries in developing effective defenses, the concept of zero-trust doctrine can also be integrated into game models by establishing relevant adversarial models. By incorporating the principles of zero trust, game models can create decision-dominant zero-trust policies to defend against APTs in 5G networks.

To this end, we propose a decision-dominant zero-trust defense (DD-ZTD) against adversarial attacks in 5G networks in MDW to strike the right balance between information acquisition and fast decision-making. DD-ZTD is built on a game-theoretic framework that captures the information asymmetry and the competitive nature of cyber defense. Following the ``never trust, always verify'' principle \cite{rose2020zero}, zero-trust defense (ZTD) equips the defender with a proactive information processing mechanism when operating with incomplete information about the attacker's intentions, capabilities, and actions, which is crucial to develop strategies that account for the information asymmetry.  

{The ZTD problem of the 5G network is modeled as an \textit{asymmetric information Markov game} (AIMG) between the defender and the attacker. Thanks to its great expressivity, AIMG offers a comprehensive characterization of various information structures in cyber defense, which facilitates defense design in various security contexts. Furthermore, the equilibrium notion in AIMG lays a theoretical underpinning of an adaptive ZTD in the presence of information asymmetry. Powered by recent advancements in machine learning, the proposed game-theoretic ZTD framework exhibits great potential in devising a generalizable intelligent defense against a wide range of cyber attacks arising from a variety of network systems possibly unknown to the defender beforehand.  }

To outpace the attacker in the cyber kill chain, ZTD is further augmented by decision dominance (DD), where DD accelerates the defense decision-making in ZTD. As its name suggests, DD makes the defender the dominant player in the dynamic game by taking decisive actions based on acquired partial information with high confidence before the attacker compromises the network system, sharing the same spirit of the motto ``first look, first shot, first kill.''\cite{Osborn_2018}. Such strategic dominance is achieved by game-theoretic calculations where the defender takes into account the attacker's decision-making process. {DD amounts to an optimal stopping (Dynkin's) game problem, which essentially captures the defender's strategic anticipation of the opponent's stopping criterion, as well as the fundamental tradeoff between the benefits and harm of lingering in the interaction, which is ubiquitous in the cyber security domain. The equilibrium notion for DD enables the defender to make opponent-independent stopping decisions based on the payoff evaluation for the underlying cyber kill chain process while making the monitoring and investigation as effective as possible.}

The rest of this chapter is organized as follows. Section~\ref{sec:mdw} provides an overview of multi-domain warfare and associated 5G networks across multiple domains, laying the context for further discussions. Section~\ref{sec:attack} articulates the emerging security challenges in 5G networks, particularly the advanced persistent threats (APT). To address these security issues, we propose a decision-dominant zero-trust defense for 5G networks in Section~\ref{sec:game}, where the game-theoretic conceptualization is presented. Section~\ref{sec:ztd} and Section~\ref{sec:dd} dive into the details of the zero-trust defense and the decision-dominance concept in detail, respectively, where case studies of the proposed DD-ZTD are presented. 
}

{
\section{Multi-Domain Warfare and 5G Networks}
\label{sec:mdw}
This section briefly overviews multi-domain warfare and the associated 5G communication networks.
\subsection{Multi-domain Warfare}
\label{subsec:mdw}
Multi-domain warfare (MDW), a new operation concept designated by the U.S. Army \cite{gady2020cyber},  refers to the combined arms employment of military capabilities straddling multiple domains to create and exploit a decisive advantage over an adversary. Unlike traditional warfare, where operations are conducted within a single domain, MDW rests on synthesizing various military capabilities across five warfighting domains: land, sea, air, space, and cyberspace. 

The backbone of MDW is the coordination and integration among different military units from multiple domains, leading to joint operations where various military services, such as the army, navy, air force, and space force, work together collaboratively. By operating across multiple domains, military forces can disrupt an adversary's operations and degrade their ability to fight.

\subsection{5G Multi-Domain Networks}
\label{subsec:mdn}
One challenge to achieving real-time coordination and integration in multi-domain warfare is the lack of network infrastructure to support interoperability among military units using different communication systems, making coordinating actions across multiple domains difficult. The fifth generation (5G) wireless communication technology plays a vital role in multi-domain warfare. It provides a network infrastructure that enables faster data transfer speeds, greater bandwidth, lower latency, and increased capacity and reliability than previous generations of mobile networks. Thanks to its advanced features, 5G technology provides the foundation for faster, more connected, and more capable military operations across multiple domains, leading to improved situational awareness, enhanced command and control, precise targeting, integration of unmanned systems, and support for emerging technologies like the internet of battlefield things(IoBT). We elaborate on these aspects in the ensuing paragraphs. Figure~\ref{fig:5g} presents a schematic illustration.

\runinhead{Situational Awareness} 5G MDN can support the transmission of large volumes of data in real time. This enables the rapid exchange of information between sensors, platforms, and command centers across different domains. Improved situational awareness allows military commanders to make more informed decisions and respond promptly to changing battlefield conditions.

\runinhead{Precise Targeting} The low latency and high bandwidth of 5G networks enable the real-time transmission of sensor data and imagery, supporting the precise targeting of enemy assets. This enhances the effectiveness of kinetic operations, such as precision strikes, and improves the accuracy of intelligence, surveillance, and reconnaissance (ISR) capabilities.

\runinhead{Command and Control} 5G networks can facilitate seamless communication and coordination between military units and commanders across domains. Reliable and low-latency connectivity enables the transmission of commands, orders, and mission-critical data, enhancing command and control capabilities in multi-domain operations.

\runinhead{Integration of Unmanned Systems and IoBT} 5G connectivity can facilitate the communication and control of unmanned systems and autonomous vehicles, both on the ground and in the air. This enables the integration of unmanned assets into multi-domain operations, enhancing their situational awareness, coordination, and responsiveness. In addition, 5G connections among a massive number of devices and sensors can be leveraged to create a comprehensive network of interconnected assets. This integration allows for better monitoring, management, and control of unmanned systems, autonomous vehicles, and other IoT devices across domains.
}

{
\section{Emerging Security Challenges in 5G Multi-Domain Networks}
\label{sec:attack}

5G networks represent a significant advancement in technology, offering functionalities that set them apart from previous generations. In the context of multi-domain warfare, it is crucial to examine the vulnerabilities inherent in 5G networks, as they can be exploited to form an APT kill chain. This section will delve into the vulnerabilities stemming from APIs, network slicing, and the supply chain.
 
\subsection{Security of 5G Multi-Domain Networks}

5G networks play an important role in MDW as they provide a network infrastructure that enables faster communication, greater bandwidth, and lower latency between different military units compared to previous generations of mobile networks.   With 5G technology, military personnel can access and share information in real-time, allowing for faster decision-making and more efficient deployment of resources. For example, a military unit is conducting a mission in an urban environment that involves ground troops, drones, and surveillance equipment. The troops on the ground need to communicate with each other in real time while also receiving information from the drones and surveillance equipment to coordinate their actions.

Moreover, 5G technology allows for the use of advanced technologies such as drones, autonomous vehicles, and augmented reality, which can be used to gather intelligence, conduct surveillance, and engage in combat operations. These technologies rely on high-speed, low-latency networks to function effectively, and 5G provides the necessary infrastructure to support their deployment. For example, during the U.S. military's operations in Afghanistan, the 5G-satellite communication network was used to provide real-time communication and intelligence sharing between ground forces, aircraft, and command centers. The system enabled military forces to coordinate their actions across different domains while also providing them with the information and intelligence needed to make informed decisions.
 
In addition to its communication capabilities, 5G-supported satellite networks also have the ability to support other mission-critical functions, such as intelligence gathering and surveillance. The system's high-capacity communication services and advanced technology make it a critical enabler for multi-domain warfare, providing military forces with the network infrastructure needed to support real-time communication and information sharing across different domains.

Recent years have seen that the adoption and implementation of 5G networks for military applications are gaining momentum.  As military forces become more reliant on 5G networks, they also become more vulnerable to cyber-attacks. To achieve multi-domain warfare, military forces need to develop robust cybersecurity measures to protect their 5G networks and systems from cyber threats. One critical threat is APT attacks on 5G networks. APT attacks are typically carried out by skilled and well-funded attackers who use sophisticated techniques to gain unauthorized access to sensitive information and systems. APT attackers may conduct extensive network reconnaissance to gather information about the 5G network and its vulnerabilities. They exploit vulnerabilities in the 5G network and gain unauthorized access to a device or system within the network to move laterally through the network and access other devices or systems within it. In 5G networks, lateral movement capabilities can be particularly dangerous, as they can allow attackers to gain access to critical systems and data within the network. For example, an attacker who gains access to a single device within a 5G network could potentially use lateral movement techniques to gain access to other devices or systems, such as servers or databases, which contain sensitive or confidential data.
 
%(MORE ABOUT APT CAN BE ADDED.)

\subsection{5G Threat Landscape: Vulnerabilities and Kill Chain}

The emergence of 5G technology represents a significant departure from previous mobile generations, bringing with it a distinct set of security requirements. This is particularly crucial for military users who often necessitate tailored and specialized services to address their unique operational needs.
There are several key threats associated with 5G networks beyond general cybersecurity threats (e.g., unauthorized access, human errors, and misconfigurations). Various threat frameworks are available to aid in analyzing these threats, such as those provided by MITRE Fight and 3GPP's Security Assurance Specifications (SCAS) and Technical Specification (TS) 33.501. 

One prominent threat to 5G networks is virtualization threats, which impact virtual machine (VM) and container service platforms, affecting various aspects of 5G, including the Core, RAN, MEC, Network Slicing, Virtualization, and Orchestration and Management. These threats encompass DoS attacks, VM/container escape, side-channel attacks, and misconfigurations by cloud service consumers. For instance, extreme resource consumption by one tenant in a multi-tenant virtualization environment can lead to a DoS event for neighboring tenant systems, impeding mission functionality. Similarly, colocation attacks, such as VM/container escape or side-channel attacks, can compromise neighboring compute workloads, resulting in resource deprivation, lateral movement, and compromising data confidentiality, integrity, or availability. A side-channel attack on 5G RAN or Core functions could allow bypassing user account permissions, virtualization boundaries, or protected memory regions, thereby exposing sensitive information.

One type of threats is on 5G network slices. These threats may exploit weaknesses in the network slice's configuration, protocols, or applications, potentially leading to unauthorized access, data breaches, or service disruptions within that particular slice. To combat this threat, slice isolation is a promising approach. It involves creating and maintaining separate virtual network slices within the 5G infrastructure. By isolating slices, potential interference or vulnerabilities in one slice are contained, ensuring the integrity and security of other slices. 

% https://www.networkworld.com/article/3697269/5g-network-slices-could-be-vulnerable-to-attack-researchers-say.html

 %https://www.bleepingcomputer.com/news/security/nsa-shares-tips-on-mitigating-5g-network-slicing-threats/

As 5G networks utilize application programming interfaces (APIs) for communication and interaction between different components, several potential threats can arise. These include DoS attacks targeting 5G APIs by overloading them with a high volume of requests or exploiting API vulnerabilities to exhaust system resources. Attackers can also exploit API vulnerabilities by abusing or misusing them to gain unauthorized access, manipulate data, or disrupt services. This can involve sending malicious API requests, performing injection attacks, or overwhelming the API with excessive requests (API flooding).

%https://arstechnica.com/information-technology/2022/08/one-of-5gs-biggest-features-is-a-security-minefield/?comments=1&comments-page=1

%https://cradlepoint.com/resources/blog/5g-iot-security-network-slicing-ztna-and-other-things-you-need-to-know/

The increasing complexity of 5G networks involves a vast ecosystem of suppliers and vendors. Security vulnerabilities in the supply chain can lead to compromised components or malicious software being introduced into the network infrastructure, posing significant risks. For example, the presence of counterfeit or substandard components in the 5G supply chain poses significant risks to network security and integrity. These components may not meet the required quality standards or security specifications, making them susceptible to exploitation and compromise. Unauthorized actors could exploit these vulnerabilities to gain unauthorized access or control over the network infrastructure, potentially leading to data breaches, service disruptions, or unauthorized surveillance.

% https://www.softeq.com/blog/how-to-ensure-5g-supply-chain-security
% https://www.cpomagazine.com/cyber-security/securing-the-supply-chain-of-the-5g-network-is-critical-to-its-success/

In addition to counterfeit components, there is a risk of introducing malicious software or hardware into the 5G supply chain. This can occur through intentional modifications or the inclusion of backdoors that provide unauthorized access points. Threat actors can exploit these vulnerabilities to infiltrate the network infrastructure, compromise the confidentiality, integrity, and availability of data, or gain unauthorized control over critical network functions.

Supply chain security risks can also originate from third-party providers involved in the network deployment, such as installation contractors or maintenance service providers. Inadequate security measures implemented by these third parties, insider threats, or the compromise of their systems can introduce vulnerabilities into the 5G network. Weaknesses in the security practices of these entities can be exploited by threat actors, compromising the overall security of the network.

%These threats expose a significant attack surface that attackers can exploit to their advantage. Attackers can orchestrate a kill chain, which involves a series of stages and techniques, with each step building upon the previous one. For instance, they may initiate the attack by leveraging social engineering tactics to gain initial access to the network, then exploit zero-day vulnerabilities present in 5G networks. With escalated privileges, they can move laterally across the 5G network, traversing from UE devices to core networks, and ultimately targeting specific assets.

 \begin{figure}[!ht]
    \centering
    \includegraphics[width=0.8\textwidth]{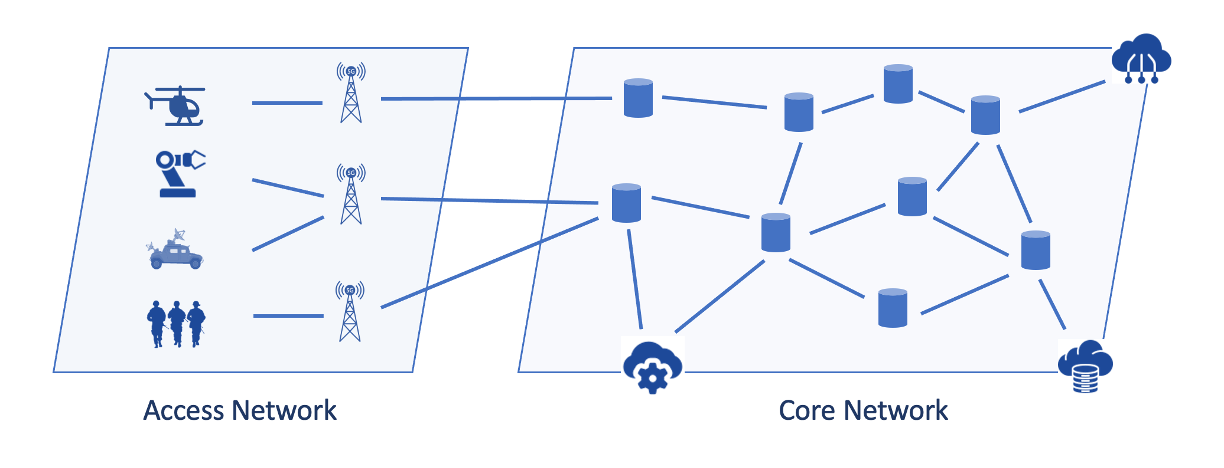}
    \caption{An illustration of 5G network consisting of access network and core network as two major components. The core network has the functionalities of control plane and user plane separation, network functions virtualization (NFV), network slicing, mobility management, and multi-access Edge Computing (MEC)}
    \label{fig:5gnet}
\end{figure}

The combination of vulnerabilities in API, supply chain, and network slicing, along with others, can be exploited by an Advanced Persistent Threat (APT) attack to form a comprehensive kill chain. Fig. \ref{fig:5gnet} provides a visual representation of a baseline 5G network, where UEs utilizing O-RAN technology connect to the 5G core networks. This interconnected infrastructure presents an attack surface that an adversary can leverage to target specific entities.
By capitalizing on the identified vulnerabilities, an attacker can exploit weaknesses in the API layer, infiltrate compromised components introduced through the supply chain, and exploit insufficient isolation or monitoring within the network slicing architecture. This enables the attacker to establish a persistent presence within the network and navigate through various stages of the kill chain to reach their intended target. Fig. \ref{fig:5g} has illustrated the potential attack path an adversary may take, highlighting the entry points, lateral movement, and potential impact on the 5G network. Understanding and visualizing this attack surface assists in identifying critical areas for security enhancements and mitigations.

 Zero-trust policies can be implemented to counteract such threats. It aims to establish clear rules and guidelines for access, authentication, and data protection within the network. These policies define which individuals or entities have access to specific resources, under what conditions, and the level of authorization required. It is crucial for the policy to align with the organization's security objectives and regulatory requirements. Regular monitoring of network traffic, user behavior, and access logs is essential to promptly identify any anomalies or potential security breaches. Additionally, it is important to periodically review and update the Zero Trust policy to adapt to evolving threats and changes in the network environment. 

}

{
\section{Decision-Dominant Zero-Trust Defense: A Game-Theoretic Framework}
\label{sec:game}
This section presents a high-level overview of the proposed decision-dominant zero-trust defense (DD-ZTD) in 5G multi-domain networks, arguing that the proposed game-theoretic framework leads to a unified framework for cyber defense in 5G networks. 

\subsection{Decision Dominance}

Decision dominance refers to the ability of a defender to outmaneuver the adversary's decision-making cycle by possessing superior situational awareness and efficient reasoning capabilities. It involves making timely and effective decisions, particularly in unknown and uncertain situations, in order to gain an advantage over the attacker. To achieve decision dominance, a defense strategy needs to excel in two stages: monitoring and detection and response. In the monitoring and detection stage, a decision-dominant defense can gather, process, and analyze information from various sources to obtain a comprehensive understanding of the cyber operational environment. This enables the defender to proactively identify and assess potential threats. In the response stage, a decision-dominant defense can swiftly evaluate available options, assess risks, and make informed decisions in a timely manner. By doing so, it can effectively thwart planned attacks before they are executed. Achieving decision dominance requires proactive cyber mechanisms like cyber deception and attack engagement to gather immediate intelligence. Agility is also crucial, allowing the defender to learn, adapt, and respond to changing situations, seize opportunities, and adjust strategies and tactics as necessary.

Zero-trust decision-dominance strategies refer to a specific type of decision-dominance strategy that operates on the assumption of the presence of adversaries at all times. These strategies are particularly critical for securing 5G networks, given the expanding attack surface and the significant number of IoT devices deployed in battlefield environments. Implementing these strategies requires strategic thinking and continuous monitoring of device behaviors to assess their trustworthiness. Timely evaluation and rapid response capabilities are essential in terms of network configuration and access control policies to counteract adversaries before they can execute their planned attacks. To ensure effective implementation, it is necessary to establish quantitative and formal frameworks that incorporate zero-trust decision-dominance into 5G network security policies. These frameworks provide a structured approach to design and enforce robust security measures that align with the principles of zero trust, enhancing the overall resilience and protection of 5G networks in dynamic threat environments.

\subsection{Conceptualization of Decision-Dominant Zero-Trust Defense}
\label{subsec:overview}

One of the primary objectives of this book chapter is to develop a quantitative framework that formalizes the decision-making process for zero-trust defense. The inherent competition between attackers and defenders naturally gives rise to a dynamic game environment that reflects the win-lose nature of multi-stage interactions. To account for the information asymmetry between the players resulting from differences in monitoring and sensing capabilities, we propose a dynamic game of asymmetric information. In this game, players utilize the information available to them through the established information structure to infer unknowns. Variations in the information structure lead to differing belief structures. Players make decisions based on their beliefs, resulting in new observations in subsequent rounds of interaction and the formation of updated beliefs. It is evident that there exists interdependence between the beliefs and actions arising from the players' chosen strategies. The solution concept for the game necessitates consistency between the agents' beliefs and their optimal effort strategies. This concept gives rise to the notion of Bayesian Nash equilibrium, which serves as the foundation for developing algorithms to implement game-theoretic solutions in practical scenarios.

It is important to note that belief formation stems from incomplete information regarding the other agent. In our case, the incomplete information pertains to the behavior of the other player. Thus, it can also be seen as a process of establishing trust in the other player. This naturally aligns with the concept of zero trust, which requires the defender to distrust users or third-party players in the network despite their credentials. At the outset, the true identity must be considered unknown and untrusted, and the evaluation of a player's trustworthiness epitomizes the principle of zero trust. The baseline equilibrium concept is established using Bayesian rationality, where Bayes' law is employed to update beliefs whenever new observations are obtained by the players. In practice, this baseline can be replaced with a machine-learning approach for inference. In modern scenarios, vast amounts of data are collected from numerous users interacting with the system. These data can be incorporated into game-theoretic models, facilitating the practical application of equilibrium solution concepts. Detailed models and their applications to lateral movements will be discussed in the subsequent section.

In order to accommodate the requirement of quick decision-making in decision-dominant scenarios, the game becomes dynamic and no longer has a fixed horizon. In this type of game, known as a stopping time game, players have the ability to choose when to cease observations and make their decision. The advantage of stopping early lies in determining the payoffs, but there is a risk of uncertainties that may lead to higher payoffs if the decision is postponed. However, it is important to note that the other player also has the capability to terminate the game. If the attacker terminates the game prematurely, the defender would be in a passive position. Thus, the competitive nature of the game naturally leads to a decision-dominant scenario. The defender's reasoning involves inferring the opponent's strategies based on the observations and, in the meantime, trades off between the probable stopping by the attacker as well as the low payoff as a result of early stopping.  To formally capture this dynamic, we introduce a stopping-time game in the ensuing section, with the aim of creating decision-dominant strategies. The associated Nash equilibrium solution concept allows us to reason formally about the active and passive situations of the defender, referred to as defender dominance and adversary dominance, respectively. The baseline analysis provides insights into the necessary structures for developing winning solutions, including the payoff structures, information structures, and inference mechanisms. This analysis also establishes a theoretical foundation for understanding the fundamental limits of strategic decision dominance in the face of a strategic adversary. By integrating decision-dominant strategies with zero-trust defense strategies within the baseline framework, we can establish a symbiotic relationship between the two. Additionally, the consolidation and integration of data analytics can pave the way for the development of practical algorithms in the future.

%Can we have a time varying information structure? The information structure is controlled or instructed by knowledge?
%We can also create an information acquisition game for this purpose. In this game, the agents also compete for gathering the data. It will be related to the framework by Yunhan.

%\textcolor{blue}{In this section, we create a class of game-theoretical models to capture ...(SOME OVERVIEW REMARKS. ALSO LITERATURE)}

%LITERATURE: 
The proposed framework in this book chapter is solidly built on the recent development of game-theoretic models for cybersecurity. Recent advances have witnessed the growth in their application to assess security risks, design protection mechanisms, and inform policy making for communication networks \cite{mallik2000analysis,mukherjee2012jamming,sayin2018game}, Internet of things \cite{Chen2019optimal,pawlick2015flip,pawlick2017strategic}, power and energy systems \cite{huang-gamesec-17,chen2022cross,chen2019game,chen2016game}, manufacturing and robotics \cite{chen-spie-19,chen-TCNS-19-games,chen-CDC-16,zhu2021cybersecurity}, supply chains \cite{kieras2020riots,ge2022accountability,kieras2022iot}, and transportation networks \cite{pan2022poisoned, pan2023resilience, pan2023stochastic}. Game theory has also provided theoretical foundations for cyber deception \cite{zheng2012dynamic,zhu2012deceptive,zhuang2010modeling,pawlick2019game}, moving target defense \cite{zhu2013game,jajodia2011moving}, and human behaviors \cite{huang2022radams,huang2021combating,huang2023cognitive}. Both decision-dominance and zero-trust defense possess distinct characteristics that necessitate specific game structures to capture their essential features and provide valuable insights. In this context, our focus lies on two types of game structures: the game of asymmetric information and stopping time games. This chapter not only applies these game structures to 5G zero-trust security problems but also contributes to a novel class of game-theoretic frameworks, pushing the boundaries of game theory forward. 

Our contribution primarily revolves around the creation and analysis of stopping-time games within the framework of asymmetric information dynamic games. By incorporating asymmetric information into these games, we introduce a new dimension that enhances our understanding of strategic interactions. Furthermore, we consolidate the fields of meta-learning and explainable learning within the domain of asymmetric information games, fostering a comprehensive approach to game analysis. Through these contributions, we aim to extend the frontiers of game theory, providing researchers and practitioners with valuable tools to tackle decision-dominance and zero-trust defense challenges effectively.

}

{
\section{Zero-Trust Defense }
\label{sec:ztd}
With a growing threat landscape and attack surfaces in 5G networks, traditional perimeter-based defense, a static defense mechanism, has become inadequate in the face of sophisticated cyber attacks, such as APTs. Advanced attackers can evade traditional intrusion detection at the perimeter, obtain privileges as an insider with stolen credentials, and move laterally within the network. In response to the vulnerabilities in the static defense,  zero trust emerges as a promising security framework, assuming that no entities can be trusted and therefore requiring verification processes for every incoming access request \cite{rose2020zero}. 

Zero-trust defense (ZTD) consists of two components: trust evaluation and access policy. Square one of ZTD is to quantitatively establish the trustworthiness of each entity in the network, which is highly nontrivial in 5G networks with large-scale heterogeneous network entities. Due to the increasing network connectivity, the defender can only acquire limited partial observations of the user's trace through methods such as Intrusion Detection Systems \cite{LIAO201316}, and Security Information and Event Management \cite{siem}. These limited observations create \textbf{information asymmetry}, complicating the defender's decision-making, and a quantitative metric measuring the user's trustworthiness using partial observations is indispensable.    

With the trust evaluation, the defender can enforce different policies for access to network resources. What distinguishes ZTD from the perimeter-based one is that the trust evaluation and the access policy, together with the network monitoring unit, constitute a feedback loop shown in Figure~\ref{fig:feedback}.  As new observations are fed into the evaluation unit, the defender adjusts the trust and the access policy accordingly, leading to a dynamic defense. This section articulates a game-theoretic framework (see \Cref{def:mg}) for ZTD design in 5G networks, which offers a natural set of tools to capture the information asymmetry and the competitive nature of the two parties in dynamic environments. 
\begin{figure}
    \centering
    \includegraphics[width=\textwidth]{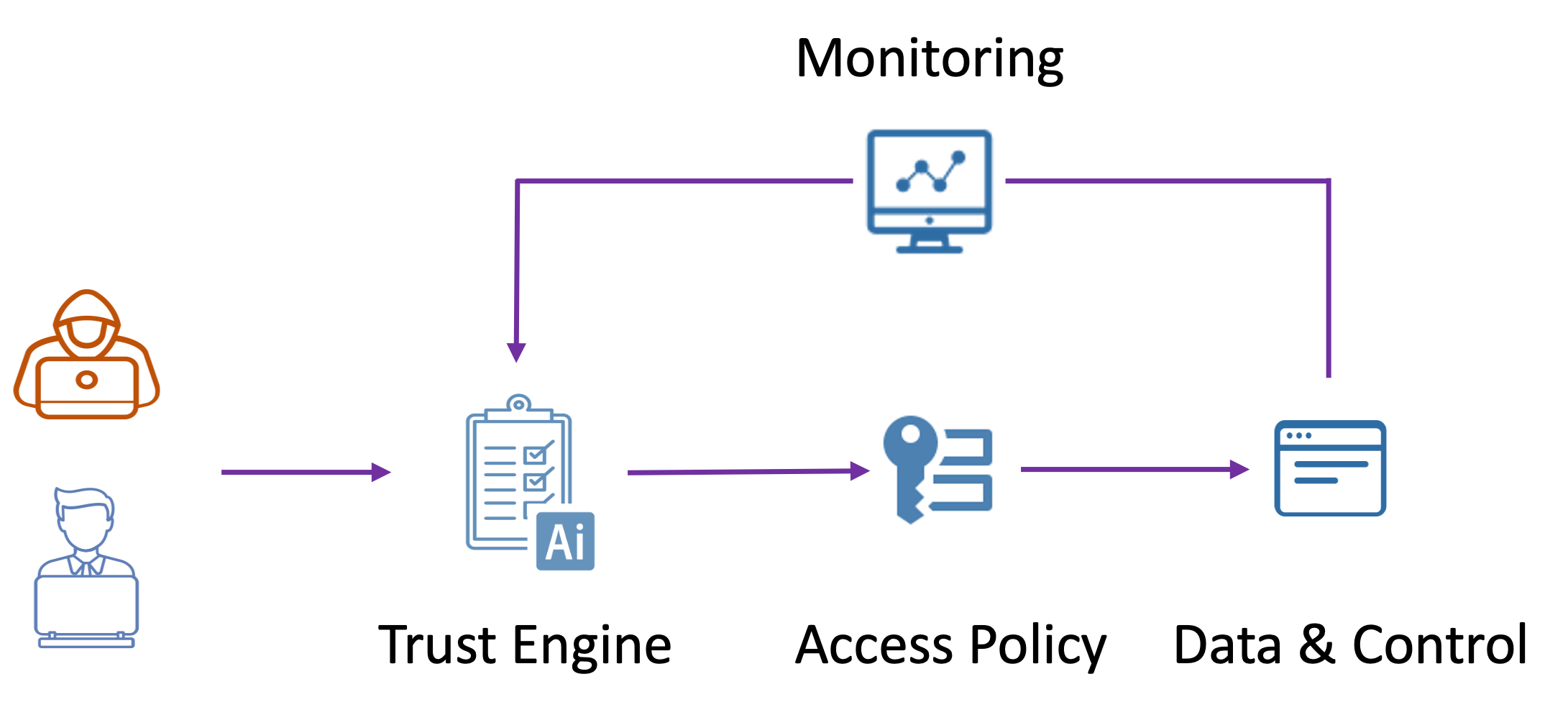}
    \caption{An illustration of the feedback loop in zero-trust defense (ZTD) architecture. Unlike the perimeter-based defense, ZTD dynamically evaluates the trustworthiness of the user using feedback from the security monitoring system, such as SIEM \cite{siem}. Based on the trust evaluation, the access policy either grant or deny access}
    \label{fig:feedback}
\end{figure}

The proposed game-theoretic framework provides a theoretical underpinning of adaptive and strategic ZTD built upon the notion of perfect Bayesian Nash equilibrium (see \Cref{def:pbne}) in the face of asymmetric information. This equilibrium-based ZTD can be further augmented with modern machine-learning (ML) methodologies providing an end-to-end automated network defense (see \Cref{subsec:trust and policy}), generalizing to adversarial scenarios unseen in the pre-training stage. As advanced ML machinery enters the picture, the ZTD architecture grows opaque to human operators. To make ML-based ZTD itself trustworthy to humans, it is necessary to increase the explainability and accountability of learning-based ZTD, which is discussed at the end of this section.

\subsection{Information Asymmetry in Zero-Trust Defense}
As a prevailing phenomenon in security applications \cite{tao_info}, information asymmetry refers to the fact that one party is better informed than the other party at the point of decision-making. To facilitate our discussion, we use the notion \textbf{information structure} \cite{tao_info} to capture the player's observations and knowledge throughout the decision-making process, which is mathematically a set of random variables whose realizations can be observed by the player \cite{tao_info}. We first present a bird's eye view of asymmetric information structures in the cyber defense of 5G networks, and mathematical definitions and arguments are deferred to Definition~\ref{def:mg} and the ensuing remarks. 

Compared to its predecessors, 5G networks enjoy increasing capacity and reliability that can support a massive number of heterogeneous devices. Consequently, it becomes prohibitive, if not impossible, for either the defender or the attacker to acquire a holistic view of the underlying network. The resulting information structures of both parties' partial observations display complexities to various extents, which can be categorized according to different taxonomies. We here present two taxonomies based on the notion of information superiority proposed in \cite{tao_info}: one player is said to be informationally superior to the other if its information structure is a superset of its counterpart.  

Depending on which party acquires the information superiority, information asymmetry includes \textbf{one-sided} and \textbf{double-sided} information asymmetry. One-sided information asymmetry refers to a situation where one party achieves information superiority over the other. If no one is informationally superior, then the resulting situation is of double-sided information asymmetry, where both parties acquire private information hidden from the other \cite{li2022commitment}.   

Depending on whether the information superiority is rooted in the knowledge or the observation, information structures can be categorized into \textbf{incomplete} and \textbf{imperfect} information structures. Knowledge is endogenous, reflecting the player's comprehension of the decision-making process. The incomplete information points to the player's uncertainty regarding the other's decision-making capabilities and incentives.  In contrast, observation is exogenous, referring to the player's awareness of events that have previously occurred. Imperfect information refers to the situation where the player is unaware of some events in the decision-making.

As one shall see later in the running example in \Cref{subsec:lateral}, the aforementioned information structures are prevalent in network defense. To systematically investigate information asymmetry in the cyber defense of 5G networks, we propose the asymmetric information dynamic games in the following, laying a mathematical foundation to facilitate ZTD design under sophisticated information structures, which is visualized in \Cref{fig:aimg}.
\begin{figure}
    \centering
    \includegraphics[width=\textwidth]{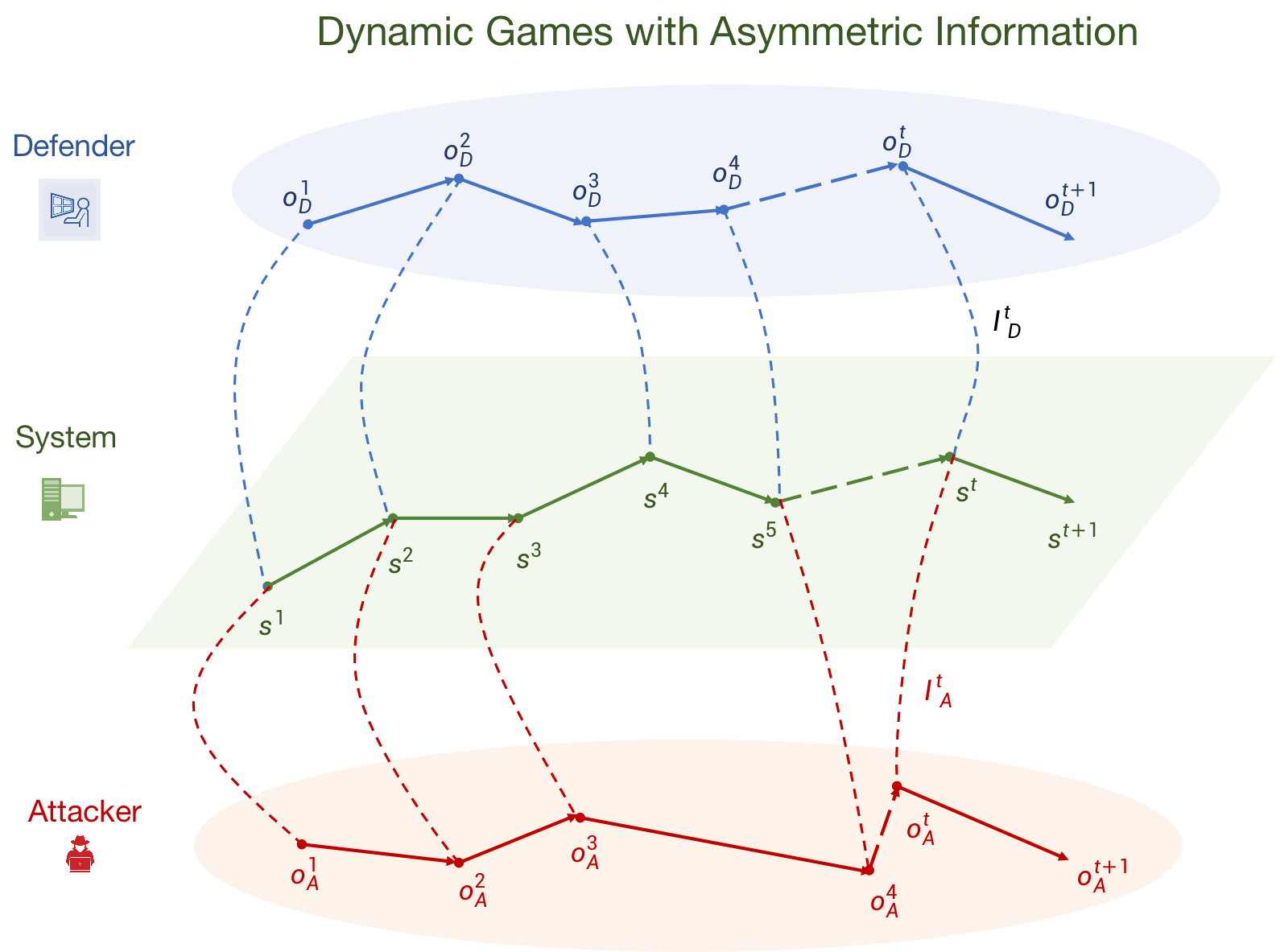}
    \caption{An illustration of asymmetric information dynamic games defined in \Cref{def:mg}. Let $D$ and $A$ denote the defender and the attacker, respectively. Under asymmetric information structures ${I}^t_D$ and $I_A^t$, the two players have disparate partial observations, denoted by $o^t_D, o_A^t $ on the system operation $s^t$. In zero-trust defense, the defender must infer the attacker's intention, assign trust scores, and determine the access policy based on its limited observations, which calls for efficient and adaptive trust evaluation and policy learning}
    \label{fig:aimg}
\end{figure}
\begin{svgraybox}
\begin{definition}[Asymmetric-Information Markov Game]
\label{def:mg}
    An asymmetric-information Markov game (AIMG) $\mathcal{G}$ is given by the following tuple 
    $$\mathcal{G}:=\langle \mathcal{N}, \Omega, \rho, \mathcal{S}, (\mathcal{O}_i)_{i\in \mathcal{N}}, ( \mathcal{A}_{i})_{i\in \mathcal{N}}, P, (u_i)_{i\in \mathcal{N}}, (\sigma_i)_{i\in \mathcal{N}}, ({I}_i)_{i\in \mathcal{N}}, H\rangle,$$
    where the definition of each component within the tuple is as below. It is assumed every set is discrete and finite. Let $t\in \mathbb{N}_{+}$ be the time index.
    \begin{itemize}
        \item $\mathcal{N}=\{D, A\}$ is the decision-maker (player) set, including the defender and the attacker, denoted by $D$ and $A$, respectively. For simplicity, we consider a single attacker within the network, and the generalization to the case where multiple attackers coexist is straightforward. 
        \item $\Omega$ is the attacker's type space, and its typical element $\omega$ indicates its attack capability (e.g., stealthiness) and objective (e.g., data breach). To simplify the exposition, the normal user is also treated as one type of attacker without malicious intentions or attack capabilities. 
        \item $\rho$ is the type distribution over $\Omega$, and $\rho(\omega)$ implies the probability of a certain attacker $\omega$ appearing in the network.
        \item $\mathcal{S}$ denotes the state space with its typical element $s$ representing the operation status of the network. 
        \item $\mathcal{O}_i$ denotes the observation space, and its typical element $o_i$ represents the player $i$'s partial observation.
        \item $\mathcal{A}_i$ is the action space of the player $i$.
        \item $P:\mathcal{S}\times (\mathcal{A}_i)_{i\in \mathcal{N}}\times \Omega \rightarrow \Delta(\mathcal{S})$ is the state transition function, depicting how the network operation evolves under the joint force of the defense and attack. To be specific, $P(s^{t+1}|s^t,a^t_D,a^t_A, \omega)$ gives the probability that $s^{t+1}$ emerges after the two players execute $a^t_D$ and $a^t_A$ at the state $s^t$. 
        \item $u_i: \mathcal{S}\times(\mathcal{A}_i)_{i\in \mathcal{N}}\times \Omega\rightarrow \mathbb{R} $ is the instantaneous cost of the player $i$.
        \item $\sigma_i: \mathcal{S}\times(\mathcal{A}_i)_{i\in \mathcal{N}}\times \Omega\rightarrow\Delta(\mathcal{O}_i)$ is the observation function, and $\sigma_i(o^t_i|s^t, a_D^t, a_A^t, \omega)$ denotes the probability of observing $o_i^t$ when the underlying state is $s^t$.
        \item $I_i$ is a set-valued mapping, characterizing the information structure of the player $i$ throughout the Markov game.  Let $\mathcal{H}^t:=\{\omega, [s^k(a^k_i o_i^k)_{i\in \mathcal{N}}]_{k=1}^{t-1} s^{t}\}$ be the history of the gameplay up to time $t$, then $\mathcal{I}_i^t:=I_i(\mathcal{H}^t)\subset \mathcal{H}^t$ presents the player's partial observation of the play. 
        \item $H$ is a constant, denoting the horizon length of the game, i.e., the operating lifetime of the network. 
    \end{itemize}
\end{definition}
\end{svgraybox}
The AIMG unfolds as follows. In the first stage, a type-$\omega$ attacker is realized according to the distribution $\rho$, and the network state $s^1$ is initialized. At the time $t$, each player implements an action $a_i^t$ from the action set $\mathcal{A}_i$ based on the information structure $\mathcal{I}_i^t$. Then, the state evolves to $s^{t+1}$. This procedure repeats until the game reaches the end of the horizon. The goal of type-$\omega$ attacker is to find a policy $\pi_A:\mathcal{I}^t_A\rightarrow \Delta(\mathcal{A}_A)$ within a specified policy class $\Pi_A$ such that the cumulative cost is minimized:
\begin{equation}
    \min_{\pi_A\in \Pi_A}\mathbb{E}\left[\sum_{t=1}^H u_A(s^t, a^t_D, a_{A}^t, \omega)\right],
    \label{eq:attack-goal}
\end{equation}
where the expectation is taken over Borel probability measures in AIMG, including the transition $P$, the observation functions $(\sigma_i)_{i\in \mathcal{N}}$, and the policies $(\pi_i)_{i\in \mathcal{N}}$. 

The defender's objective is more involved than \eqref{eq:attack-goal} due to the lack of information on the attack type, and a generic characterization is given by \eqref{eq:defend-goal}, where the notations are in a similar vein of \eqref{eq:attack-goal}, except that the inner expectation $\mathbb{E}_{\omega\sim \mathcal{T}}(\cdot)$ is taken over the hidden type $\omega$ with respect to the defender's subjective belief $b^t\in \Delta(\Omega)$ based on the observations $\mathcal{I}_D^t$. Such a belief constitutes the defender's trust evaluation of the user, and a mathematical characterization is presented in Definition~\ref{def:trust}. 
\begin{equation}
    \min_{\pi_D\in \Pi_D} \mathbb{E}\left\{\sum_{t=1}^H \mathbb{E}_{\omega\sim b^t}[u_D(s^t, a_D^t, a_A^t, \omega)]\right\}.
    \label{eq:defend-goal}
\end{equation}
% Mathematically, $\mathcal{B}$ is a set of probability measures $\mathcal{B}=\{b^1, b^2, \ldots, b^H\}$
\begin{svgraybox}
\begin{definition}[Trust and Trust Engine]
    \label{def:trust}
    The trustworthiness of the user at time $t$ is defined as a probability measure over the type space $b^t\in \Delta(\Omega)$, which is determined by the defender's trust engine $\Phi$ that maps the information structure $\mathcal{I}_i^t$ to the trustworthiness $b^t=\Phi(\mathcal{I}_i^t)$. The set of beliefs $\{b^t\}_{t=1}^H\in \Delta(\Omega)^H$ is referred to as the trust evaluation.
\end{definition}
\end{svgraybox}

The trust metric $b$ we consider is a probability measure, and $b(\omega)$ depicts the defender's subjective belief over the hidden type $\omega$, also referred to as the trust score \cite{tao23ztd}. With the trust evaluation, the defender can determine the access policy $\pi_D(\mathcal{I}_i^t, b^t)$ based on its observation, which, together with the trust engine, constitutes a zero-trust defense mechanism. A mathematical definition is given below. 
\begin{svgraybox}
\begin{definition}[Zero-Trust Defense]
    \label{def:ztd}
    The zero-trust defense is defined as a pair of the trust engine $\Phi:\cup_{t=1}^H\{\mathcal{I}_i^t\}\rightarrow\Delta(\Omega)$ and the access policy $\pi_D: \cup_{t=1}^H\{\mathcal{I}_i^t\}\times\Delta(\Omega)\rightarrow \Delta(\mathcal{A}_D)$. 
\end{definition}
\end{svgraybox}
Before elaborating on the two critical components of ZTD in Subsection~\ref{subsec:trust and policy}, we first remark on the expressive power of AIMG in modeling the cyber defense of 5G networks under complex information structures. In particular, Definition~\ref{def:mg} leads to a systematic characterization of various information structures, such as one/double-sided information asymmetry and incomplete/imperfect information.   
\begin{svgraybox}
\begin{definition}[One/Double-sided Information Asymmetry]

    The player $i$ is said to be informationally superior than $j$ if $\mathcal{I}_j^t\subsetneq \mathcal{I}_i^t$, for all $t$. This information asymmetry is one-sided since the player $i$ is always better informed than its opponent. If there exists $t$ such that $\mathcal{I}_i^t \setminus \mathcal{I}_j^t\neq \varnothing$ and $\mathcal{I}_j^t \setminus \mathcal{I}_i^t\neq \varnothing$, the resulting information structures are of double-sided information asymmetry. Both parties acquire private information hidden from the other, and neither achieves information superiority.
\end{definition}

\begin{definition}[Incomplete and Imperfect Information]

    For the player $i$, the AIMG is of incomplete information if $\omega\notin \mathcal{I}_i^t$ for all $t$. The AIMG is of imperfect information if there exists a $t$ such that $\mathcal{I}_i^t\setminus\{\omega\}\subsetneq \mathcal{H}^t\setminus\{\omega\}$.  
\end{definition}
\end{svgraybox}
The following uses lateral movement in 5G networks as a running example to illustrate these information structures in ZTD, which is based on \cite{tao23ztd}.
\subsection{Defending against Lateral Movement: A Running Example}
\label{subsec:lateral}
Consider a 5G network represented by a directed graph $G=\langle V, E\rangle$, where $V$ is the set of nodes, each of which represents a device/facilities connected to the network, and $E=\{(u,v)| u,v\in V\}$ denotes the set of edges, with each directed edge representing the stored service connection. For example, $(u,v)$ indicates that the user visiting node $u$ can move towards node $v$ using stored credentials. In this example, we assume that the attacker moves laterally using stolen credentials in the 5G network, attempting to reach a sensitive target node with access to some entry node such as mobile devices. The defender aims to validate the user's authentication when accessing neighboring nodes and reject the malicious attacker. This validation can be achieved by Multi-factor Authentication (MFA) \cite{ometov2018multi}. However, Each MFA over the edge incurs a cost, as MFA consumes additional security resources and time that degrade the system performance of the underlying network. The defense objective is to balance the system performance and security by strategically picking a set of edges for authentication validation.    

To demonstrate the expressive power of AIMG, we formulate the above defense problem using game-theoretic language developed in Definition~\ref{def:mg}. Two decision-makers are involved in this game: the defender and the user of an uncertain type. The user's type space is binary $\Omega=\{0, 1\}$, where $\omega=0$ indicates that the user is legitimate, whereas the user is the malicious attacker if $\omega=1$. The type distribution $\rho$ can be considered uniform since the two types are indistinguishable from the defender's viewpoint at the beginning. With historical data, the defender can treat the empirical frequency of malicious users as the type distribution, which reflects the defender's prior knowledge of the adversarial environment.  

Suppose the attacker visits a node $u$ at time $t$. Let $V^t$ be the set of neighboring nodes that can be reached using stored credentials. Mathematically, for any $v\in V^t$, there exists a $(u,v)\in E$. Denote the collection of such edges by $E^t$, and the resulting subgraph $G^t=\langle V^t, E^t\rangle \subset G$ is referred to as the authentication graph. The user can easily visit any node within the authentication graph if the defender does not impose MFA on $E^t$. Define $L^t: V^t\rightarrow \{0, 1\}$ as the indicator function. For any $v\in V^t$, $L^t(v)=1$ is $v$ has been visited before time $t$, otherwise $L^t(v)=0$. With a slight abuse of notation, we treat $L^t\in \{0,1\}^{|V^t|}$ as a binary vector of time-varying dimensions.   

The state variable comprises the authentication graph and the indicator, $s^t=(G^t, L^t)$, which captures the progress of the lateral movement and is fully observable to the attacker and the defender. With modern security machinery such as Intrusion Detection System (IDS) \cite{LIAO201316} and Security Information and Event Management (SIEM) \cite{siem}, the trace of the user/attacker creates a sequence of events that can be used for security analysis.  Consequently, the defender can acquire additional observation of the network system, which is captured by the partial observation $o_D$ in AIMG. The security machinery producing such observation corresponds to the observation function $\sigma_D$ in Definition~\ref{def:mg}. Note that the attacker's partial observation is degenerate in this case, i.e., $\mathcal{O}_A=\varnothing$. 

The action sets of the two parties are specified below. The attacker moves laterally in the network and chooses the next node to visit at each time step. Given the current state $s^t=\langle G^t, L^t \rangle$, the attacker's action set includes a collection of edges $\mathcal{A}_A:=\{(u,v)|(u,v)\in E^t, L^t(u)=1, L^t(v)=0\}$, of which the outbound node $v$ is to be visited. In APT, the stealthy attacker only picks one edge at each time step to evade detection. To combat the lateral movement, the defender strategically picks a subset of $E^t$ and imposes MFA validation accordingly. Mathematically, the defense action set amounts to the power set of $E^t$, i.e., the set of all possible subsets of $E^t$, which is denoted by $\mathcal{A}_D=2^{E^t}$.         

The system evolution is determined by the joint action of both parties, where the attacker picks an edge $a_A^t$ while the defender selects a subset of edges for MFA $a_D^t$.   Given the current authentication graph $G^t$, one needs to satisfy the MFA requirements if $a_A^t=a
_D^t$ before moving to the next node. It is assumed that the legitimate user ($\omega=0$) has a higher chance to pass this MFA, while the malicious attacker is rejected. On the occasion that $a^t_A\notin a_D^t$, both types can easily move forward. The authentication graph and the visiting history shall be updated accordingly when the user/attacker reaches a new node, and this procedure repeats until the end of the horizon. The horizon length $H\in (0, \infty)$ denotes the maximum time for the attacker to operate within the network without credential renewal. The identity life-cycle lasts for $H$ time steps, after which the stored credentials expire, and the attacker loses the foothold in the network.    

The utility function captures the trade-off between operation costs resulting from authentication and system security. From the defender's stance, the cost of authentication validation over an edge is given by the scalar $c: E\rightarrow \mathbb{R}$, and the total cost of imposing MFA on a subset of edges $a_D$ is defined as (with abuse of notation) $c(a_D)=\sum_{e\in a_D}c(e)$. In addition to the authentication cost, system security is also a key factor in the evaluation of defense effectiveness. Denote by $v^*$ the target node, and the indicator function $L^t(v^*)$ implies whether the target has been reached or not. Only when the malicious attacker ($\omega=1$) visits $v^*$, the network system is compromised, incurring a devastating cost $M$. Consequently, the defender's utility depends on the hidden type and is defined below.
\begin{equation*}
   u_D(s^t,a_D^t,a_A^t, \omega)= \left\{\begin{array}{ll}
        c(a_D^t), & \text{ if }\omega=0, \\
         c(a_D^t)+M L^t(v^*), & \text{ otherwise}.
    \end{array}\right.
\end{equation*}
Likewise, the attacker's utility function is also type-dependent. For the malicious attacker, passing the MFA is laborious and incurs a huge cost $-\hat{M}$. In contrast, the MFA validation is effortless. Whatever the type is, the attacker/user is rewarded by $R$ when arriving at the target node, and they share the same transition cost $u(s^t, a_A^t)$ when navigating within the network. Using mathematical terms, the utility function is as below.
\begin{equation*}
    u_A(s^t, a_D^t, a_A^t, \omega)=\left\{\begin{array}{ll}
       u(s^t,a_A^t)- R L^t(v^*),   & \text{ if }\omega=0,  \\
         u(s^t, a_A^t) + \hat{M} \mathds{1}_{\{a_A^t\in a_D^t\}} - R L^t(v^*), & \text{ otherwise}.
    \end{array}\right.
\end{equation*}

\subsection{Trust Evaluation and Access Policy in Zero-Trust Defense}
\label{subsec:trust and policy}
Heretofore, our discussions have primarily concerned the theoretical underpinning of ZTD provided by the game-theoretic framework (AIMG) and AIMG's expressivity regarding information structures. This subsection shifts the focus from ZTD modeling to ZTD design, and the key message is that the game-theoretic solution concept leads to effective and automated ZTD in 5G networks. 

We begin with the trust engine and trust evaluation in ZTD. Depending on its architecture, the trust engine can be categorized into attribute-based, Bayesian, and machine-learning-based trust engines. The attribute-based trust engine (ABTE) evaluates the trustworthiness of entities based on their specific attributes or characteristics. Attributes are specific properties or qualities of an entity that are relevant to determining trust, which can include factors such as the security posture of devices and endpoints, the user's location, time of access, and the sensitivity of the requested resource. The evaluation process involves assigning weights or importance to different attributes based on their significance in determining trust. These weights or importance are often pre-defined policies or algorithms, and hence, ABTE relies heavily on the domain knowledge of the security context and involves handcrafting.   

\begin{table}[!ht]
    \centering
    \caption{A comparison of three kinds of trust engines. Compared with ABTE, BTE and MLTE can adapt to new scenarios without significantly resetting the engine configuration. MLTE is a data-driven trust engine that does not require a complete grasp of the domain knowledge, yet, the price to pay is that its offline pre-training needs a decent amount of data }
    \begin{tabular}{lcccc}
    \toprule
          &  Domain Knowledge & Offline Training & Online Computation & Adaptation \\
         \midrule
         ABTE &  \cmark & \xmark & \xmark & \xmark \\
         BTE & \cmark & \xmark &\cmark & \cmark \\
         MLTE & \xmark & \cmark & \cmark/\xmark & \cmark\\
         \bottomrule
    \end{tabular}
  \label{tab:trust-engine}
\end{table}
The following subsections introduce another two trust engine architectures built upon Bayesian inference and machine learning, leading to automated dynamic trust evaluation capable of adapting to a variety of security scenarios. We refer to the two trust engines as the Bayesian trust engine (BTE) and the machine-learning-based trust engine (MLTE), respectively. A summary of these trust engines is presented in \Cref{tab:trust-engine}.

\subsubsection{Bayes Trust Engine}
\begin{svgraybox}
\begin{definition}[Bayes Trust Engine]
    A trust engine is said to be Bayesian if the trust evaluation is produced recursively using the Bayes rule. Let $\mathcal{I}_i^{t+1}\setminus\mathcal{I}_i^{t}$ be the emerging information at time $t+1$, then the trust $b^{t+1}$ is obtained by \eqref{eq:bte} and the Bayesian update is given by \eqref{eq:bayes}, where $\mathbb{P}(\mathcal{I}_i^{t+1}\setminus\mathcal{I}_i^t|\omega)$ is the probability of observing $\mathcal{I}_i^{t+1}\setminus\mathcal{I}_i^t$ conditional on the hidden type $\omega$.
    \begin{subequations}
    \begin{equation}
b^{t+1}=\Phi(\mathcal{I}_i^{t+1})=\Phi(\mathcal{I}_i^{t+1}\setminus\mathcal{I}_i^{t}, b^{t}),\label{eq:bte}
    \end{equation}
    \begin{equation}
         b^{t+1}(\omega)=\frac{b^t(\omega)\mathbb{P}(\mathcal{I}_i^{t+1}\setminus\mathcal{I}_i^t|\omega)}{\sum_{\omega'\in \Omega} b^t(\omega')\mathbb{P}(\mathcal{I}_i^{t+1}\setminus\mathcal{I}_i^t|\omega')}.\label{eq:bayes}
    \end{equation}
    \end{subequations}
\end{definition}
\end{svgraybox}

Using the lateral movement example in Section~\ref{subsec:lateral}, the emerging information for the defender at time $t+1$ is $\mathcal{I}_i^{t+1}\setminus\mathcal{I}_i^t=\{a_A^t, a_D^t, s^{t+1}, o^{t+1}\}$. Given the two parties' policies $\pi_A$ and $\pi_D$, the conditional probability is defined as $\mathbb{P}(a_A^t, a_D^t, o^t, s^{t+1}|\omega)=P(s^{t+1}|s^t, a_D^t, a_A^t, \omega)\sigma(o^{t}|s^t, a_A^t, a_D^t, \omega)\pi_D(a_D^t|s^t)\pi_A(a_A^t|s^t, \omega)$. Consequently, the belief update is obtained through the following equation. 
\begin{equation}
    b^{t+1}(\omega)= \frac{b^t(\omega)P(s^{t+1}|s^t, a_D^t, a_A^t, \omega)\sigma(o^{t}|s^t, a_A^t, a_D^t, \omega)\pi_A(a_A^t|s^t, \omega)}{\sum_{\omega'}b^t(\omega')P(s^{t+1}|s^t, a_D^t, a_A^t, \omega')\sigma(o^{t}|s^t, a_A^t, a_D^t, \omega')\pi_A(a_A^t|s^t, \omega')}. \label{eq:bayes-update}
\end{equation}

Compared with the ATE, the BTE adapts to the online environment by processing emerging information recursively without pre-training or preparation. As a plug-and-play engine, BTE requires a decent understanding of the network operation to compute the conditional probability $\mathbb{P}(\mathcal{I}_i^{t+1}\setminus\mathcal{I}_i^t)$, including the system transition $P$, the security monitoring machinery $\sigma$, and the attacker's strategy $\pi_A$. 

Several remarks are in order on the practicability of BTE. Except for the anticipated strategy $\pi_A$, the system transition function $P$ and the observation function $\sigma$ are readily accessible to the defender. In the lateral movement example, the system transition is deterministic: if one edge $(u,v)$ is picked, the next node must be the head node $v$, and the associated authentication graph and the indicator are determined accordingly. Consider the IDS as the observation function. The corresponding observation space is binary $\mathcal{O}_D=\{0, 1\}$, where $0$ means no alarm is raised while $1$ indicates that a security alert is signaled, warning the defender that the user is more likely to be malicious. In this case, $\sigma(o^t=1|s^t,a^t,a_D^t,\omega=1)$ is the detection rate, and $\sigma(o^t=1|s^t,a^t,a_D^t,\omega=0)$ is the false alarm rate, both of which are included in the IDS configuration revealed to the defender. As one can see from \eqref{eq:bayes-update}, the attacker's strategy $\pi_A$ is involved in the Bayesian update, even though it is explicitly included in the information structure $\mathcal{I}^t_D$. Due to the predictive nature of equilibrium in game theory, the defender is able to derive the attacker's optimal strategy using the game tuple in \Cref{def:mg}, from which the attacker has no incentive to deviate. Using plain words, the defender can anticipate the attacker's strategy $\pi_A$ and use this predicted strategy the update the trust. \Cref{subsubsec:policy} elaborates on this equilibrium notion in detail, where we articulate the close connection between BTE and Bayesian Nash equilibrium in game theory, leading to an adaptive zero-trust defense in contrast to ATE.         

One computational hurdle of BTE lies in that the denominator in \eqref{eq:bayes} is given by an integration (summation) of the conditional probability $\mathbb{P}(\mathcal{I}_i^{t+1}\setminus\mathcal{I}_i^t|\omega')$ with respect to the trust $b^t(\omega')$. As the arms race between the defender and the attacker heats up, the attack techniques develop day and night, and consequently, the number of attack types grows astronomical. As a result, the trust evaluation process in the online execution is burdened with great computation overhead, causing authentication latency in ZTD. 

In addition to the computation overhead, another limitation of BTE is that it relies heavily on the domain knowledge of the underlying network. Take the lateral movement defense as an example. The observation function $\sigma$ corresponds to a network security machinery (e.g., SIEM) that monitors the attacker's activities and reports incidents to network operators. Note that such feedback from the security machinery may not be directly applicable in BTE on some occasions since mathematically $\sigma$ needs to be a conditional probability measure in BTE as shown in \eqref{eq:bayes-update}.  For example, if the observation variable $o\in \mathcal{O}$ is a log message or an audit trail of the network system, then one needs to infer the attack type distribution behind these security events, requiring certain expertise in network security.

\subsubsection{Machine Learning Trust Engine}
  
To address these limitations of BTE, one alternative approach is to utilize machine learning methodologies, which offer an \textbf{end-to-end} trust evaluation. The machine-learning-based trust engine undergoes an offline training process before the online execution, and no heavy computation is involved in the online phase, although lightweight model updates can happen on some occasions to adapt the machine-learning model to new security scenarios \cite{tao23ztd}. Powered by recent advancements in large language models \cite{openai2023gpt4} and other related deep learning architectures \cite{attention, Kingma2014}, ML models capable of processing multi-modal inputs (texts and audio, etc.) display great potential in creating end-to-end trust evaluation that maps the raw system log files to a trust metric without much human involvement. Compared with BTE, MLTE does not require domain knowledge or online computation, yet the price to pay is the pre-training process, and collecting high-quality training data can be cumbersome. This is because the training data shall include incidence reports, system logs, and other related log messages, which often contain sensitive information regarding the network systems, and hence they are not open-sourced. Even if they are, these data come from a specific scenario, and the resulting trust engine may not generalize well to other network defense problems.  

Despite its limitations, MLTE provides a data-driven trust evaluation that is suitable for large-scale complex 5G networks. Mathematically, MLTE performs a statistical inference task where the engine infers the hidden type using the observations. The following takes variational Bayes inference (VB) as an example to illustrate how to train and deploy an inference network as the trust engine. In statistical inference, VB refers to a family of techniques in Bayesian inference for approximating the posterior probability of unobserved variables (e.g., hidden types) conditional on the observed ones (e.g., those in the $\mathcal{I}_i^t$). We pick VB because of its close connection with BTE and wide applications in machine learning problems, such as variational autoencoders, which gives rise to many off-the-shelf ML toolsets readily available to network security practitioners.  We refer the reader to \cite{Kingma2014} for more details on statistical inference and its applications.

For simplicity, we drop the time index in the information structure and use  $\mathcal{I}$ in the following discussion. Adopting a probabilistic viewpoint, we consider $\mathcal{I}$ and $\omega$ as two random variables generated by some random process. The process consists of two steps: 1) a realization $\omega$ is generated from the prior $\rho$; 2) a realization $\mathcal{I}$ is generated from a conditional distribution $\mathbb{P}(\mathcal{I}|\omega)$, which is in a similar vein as \eqref{eq:bayes}. The goal of the inference task is to derive the posterior distribution $\mathbb{P}(\omega|\mathcal{I})$ characterized by the Bayesian rule: $\mathbb{P}(\omega|\mathcal{I})=\mathbb{P}(\mathcal{I}|\omega)\rho(\omega)/\int \mathbb{P}(\mathcal{I}|\omega)\rho(\omega) d\omega$. Similar to the computation issue in BTE, the integral is intractable.  

Denote by $q_\phi(\omega|\mathcal{I})$ a neural network (with parameter $\phi\in \mathbb{R}^n$) approximation to the true posterior $\mathbb{P}(\omega|\mathcal{I})$. Taking inspiration from the evidence lower bound (ELBO) method \cite{Kingma2014}, we derive a loss function for the training purpose whose minimizer $q_{\phi^*}(\omega|\mathcal{I})$ serves as the trust engine in ZTD. Given a realization $\mathcal{I}$, its marginal likelihood can be written as  
\begin{equation}
    \label{eq:likelihood}
        \log \mathbb{P}(\mathcal{I})= D_{KL}[q_\phi(\omega|\mathcal{I})||\mathbb{P}(\omega|\mathcal{I})]+\mathcal{L}(\phi; \mathcal{I}),
    \end{equation}
where $\mathcal{L}(\phi; \mathcal{I})=\log \mathbb{P}(\mathcal{I}) - D_{KL}[q_\phi(\omega|\mathcal{I})||\mathbb{P}(\omega|\mathcal{I})]$. $D_{KL}[q_\phi(\omega|\mathcal{I})||\mathbb{P}(\omega|\mathcal{I})]:=\mathbb{E}_{q_\phi(\omega|\mathcal{I})}[\log( q_\phi(\omega|\mathcal{I})/ \mathbb{P}(\omega|\mathcal{I}) )]$ is the KL divergence between the two distributions. The intuition behind this likelihood expression is that the KL divergence $D_{KL}[q_\phi(\omega|\mathcal{I})||\mathbb{P}(\omega|\mathcal{I})]$ in \eqref{eq:likelihood} measures the discrepancy between the true posterior $\mathbb{P}(\omega|\mathcal{I})$ and its neural network approximation $q_\phi(\omega|\mathcal{I})$, which is to be minimized. From \eqref{eq:likelihood}, minimizing the KL term is equivalent to maximizing $\mathcal{L}(\phi; \mathcal{I})$. Since the KL term is non-negative, $\mathcal{L}(\phi; \mathcal{I})$ lower bounds the log-likelihood on the left-hand side, which is referred to as the evidence (or variational) lower bound.

Compared with the KL term $D_{KL}[q_\phi(\omega|\mathcal{I})||\mathbb{P}(\omega|\mathcal{I})]$,  this lower bound, rewritten as below, does not explicitly involve the posterior distribution $\mathbb{P}(\omega|\mathcal{I})$. The rest of this subsection is devoted to the stochastic optimization problem $\max_{\phi} \mathcal{L}(\phi; \mathcal{I})$, which amounts to the pre-training of MLTE.    
 \begin{align}
      \mathcal{L}(\phi; \mathcal{I})&=\log \mathbb{P}(\mathcal{I}) - D_{KL}[q_\phi(\omega|\mathcal{I})||\mathbb{P}(\omega|\mathcal{I})]\nonumber \\
      &=\log \mathbb{P}(\mathcal{I}) - \mathbb{E}_{q_\phi(\omega|\mathcal{I})}[\log q_\phi(\omega|\mathcal{I}) - \log \mathbb{P}(\omega|\mathcal{I}) ]\nonumber \\
      &= \mathbb{E}_{q_\phi(\omega|\mathcal{I})}[\log \mathbb{P}(\mathcal{I})]-\mathbb{E}_{q_\phi(\omega|\mathcal{I})}[\log q_\phi(\omega|\mathcal{I}) - \log \mathbb{P}(\omega|\mathcal{I}) ]\nonumber \\
      & = \mathbb{E}_{q_\phi(\omega|\mathcal{I})}[-\log q_\phi(\omega|\mathcal{I})+\log \mathbb{P}(\mathcal{I},\omega)].\label{eq:elbo}
 \end{align}   

Consider some dataset $\mathcal{D}:=\{\mathcal{I}^{(k)}\}_{k=1}^K$ consisting of $K$ independently identically distributed (i.i.d.) sample observations under random attack types $\omega^{(k)}$ drew from $\rho(\cdot)$. $\mathcal{I}^{(k)}$ represents historical security incidence reports during the network operation, and the superscript $(k)$ denotes the sample index rather than the time step. Note that only the dataset $\mathcal{D}$ is available in training, whereas the variable $\omega^{(k)}$ remains hidden (the prior $\rho$ is known), as often witnessed in real-world scenarios. 

In addition to the inference network $q_\phi(\omega|\mathcal{I})$, we introduce a generative network $p_\theta(\mathcal{I}|\omega)$, $\theta\in \mathbb{R}^m$, which approximates the conditional probability $\mathbb{P}(\mathcal{I}|\omega)$. Consequently, the joint distribution $\mathbb{P}(\mathcal{I}, \omega)$ in \eqref{eq:elbo} can also be parameterized: $\mathbb{P}(\mathcal{I}, \omega)=\rho(\omega)p_\theta(\mathcal{I}|\omega)$. With a slight abuse of notation, we denote such parameterization by $p_\theta(\mathcal{I}, \omega)$. Similar to our argument in justifying the use of $\pi_A$ in \eqref{eq:bayes-update}, $p_\theta(\mathcal{I}|\omega)$ can be interpreted as the defender's conjecture of the attack strategy that eventually leads to the resulting observation $\mathcal{I}$. With this additional parameterization, the lower bound under the datapoint $\mathcal{I}^{(k)}$ becomes 
\begin{equation}
\label{eq:bound-theta}
    \mathcal{L}(\phi, \theta;\mathcal{I}^{(k)})= \mathbb{E}_{q_\phi(\omega|\mathcal{I}^{(k)})}[-\log q_\phi(\omega|\mathcal{I}^{(k)})+\log p_\theta(\mathcal{I}^{(k)},\omega)].
\end{equation}
The remaining task is simply to approximate the gradient of the expectation in \eqref{eq:bound-theta} using samples and to apply stochastic gradient descent. Note that the expectation is taken with respect to the hidden variable $\omega$ conditional on $\mathcal{I}^k$. Hence, one needs to first draw a batch of $M$ samples $\{\omega^{(k,l)}\}_{l=1}^M$ from $q_\phi$, and then compute the gradient estimators 
\begin{subequations}
    \begin{equation}
    \label{eq:nabla-phi}
    \begin{aligned}
        \widehat{\nabla}_\phi  \mathcal{L}(\phi, \theta;\mathcal{I}^{(k)})&=-\frac{1}{M}\sum_{l=1}^K\log q_\phi(\omega^{(k,l)}|\mathcal{I}^{(k)})\nabla_\phi\log q_\phi(\omega^{(k,l)}|\mathcal{I}^{(k)}) \\
        &+\frac{1}{M}\sum_{l=1}^K \log p_\theta(\mathcal{I}^{(k)},\omega^{(k,l)})\nabla_\phi\log q_\phi(\omega^{(k,l)}|\mathcal{I}^{(k)}).
    \end{aligned}
    \end{equation}
    \begin{equation}
    \label{eq:nabla-theta}
        \widehat{\nabla}_\theta  \mathcal{L}(\phi, \theta;\mathcal{I}^{(k)})=\frac{1}{M}\sum_{l=1}^K \nabla_\theta \log p_\theta(\mathcal{I}^{(k)},\omega^{(k,l)}).
    \end{equation}
\end{subequations}
The first gradient estimation in \eqref{eq:nabla-phi} rests on a Monte Carlo (MC) estimation trick detailed below. The key message of this trick is that the gradient of an expectation can be expressed as an expectation of another gradient, which can be approximated using Monte Carlo sampling. Suppose, for the time being, one needs to estimate the gradient $\nabla_\phi \mathbb{E}_{q_\phi(\omega)}[f(\omega)]$ where $\mathcal{I}$ is suppressed, and $f(\omega)$ is an arbitrary function. Rewriting the gradient term in the integral form, we obtain
\begin{equation}
\label{eq:trick}
    \begin{aligned}
    \nabla_\phi \mathbb{E}_{q_\phi(\omega)}[f(\omega)]&=\nabla_\phi \int f(\omega)q_\phi(\omega)d\omega\\
    &=\int f(\omega) \nabla_\phi q_\phi (\omega)d\omega\\
    &=\int f(\omega)\frac{\nabla_\phi q_\phi (\omega)}{q_\phi (\omega)}q_\phi (\omega) d\omega\\
    &=\int f(\omega) \nabla_\phi \log q_\phi(\omega) q_\phi(\omega) d\omega\\
    &=\mathbb{E}_{q_\phi(\omega)}[f(\omega) \nabla_\phi \log q_\phi(\omega)].
\end{aligned}
\end{equation}
Therefore, the MC estimation under $K$ samples $\{\omega^{(l)}\}_{l=1}^K$, denoted by $\widehat{\nabla}_\phi $, is given by $\widehat{\nabla}_\phi =1/K \sum_{l=1}^K f(\omega^{(l)})\nabla_\phi \log q_\phi(\omega^{(l)})$. 

We apply this trick to derive the first gradient estimation. As one can see from the \eqref{eq:three-terms}, the gradient $\nabla_\phi \mathcal{L}(\phi, \theta;\mathcal{I}^{(k)})$ comprises three terms.  
\begin{align}
    &\nabla_\phi  \mathcal{L}(\phi, \theta;\mathcal{I}^{(k)})\nonumber \\
    &= \nabla_\phi \mathbb{E}_{q_\phi(\omega|\mathcal{I}^{(k)})}[-\log q_\phi(\omega|\mathcal{I}^{(k)})+\log p_\theta(\mathcal{I}^{(k)},\omega)] \nonumber \\
    &= \nabla_\phi \int \left( -\log q_\phi(\omega|\mathcal{I}^{(k)})+\log p_\theta(\mathcal{I}^{(k)},\omega)\right)q_\phi(\omega|\mathcal{I}^{(k)})d\omega \nonumber \\
    &=-\underbrace{\int \nabla_\phi \log q_\phi(\omega|\mathcal{I}^{(k)})q_\phi(\omega|\mathcal{I}^{(k)})d\omega}_{\text{\ding{172}}} -\underbrace{\int \log q_\phi(\omega|\mathcal{I}^{(k)})\nabla_\phi q_\phi(\omega|\mathcal{I}^{(k)})d\omega}_{\text{\ding{173}}}\nonumber \\
    &+\underbrace{\int \log p_\theta(\mathcal{I}^{(k)},\omega)\nabla_\phi q_\phi(\omega|\mathcal{I}^{(k)})d\omega}_{\text{\ding{174}}}.
    \label{eq:three-terms}
\end{align}
Since $\nabla_\phi\log q_\phi(\omega|\mathcal{I}^{(k)})= \nabla_\phi q_\phi(\omega|\mathcal{I}^{(k)})/q_\phi(\omega|\mathcal{I}^{(k)})$, $\text{\ding{172}}=\nabla_\phi\int q_\phi(\omega|\mathcal{I}^{(k)}) d\omega=0$. Applying the trick to the second and third terms, we arrive at the following equations.
\begin{align*}
    \text{\ding{173}}&=\mathbb{E}_{q_\phi(\omega|\mathcal{I}^{(k)})}[\log q_\phi(\omega|\mathcal{I}^{(k)})\nabla_\phi\log q_\phi(\omega|\mathcal{I}^{(k)}) ],\\
    \text{\ding{174}}&=\mathbb{E}_{q_\phi(\omega|\mathcal{I}^{(k)})}[ \log p_\theta(\mathcal{I}^{(k)},\omega)\nabla_\phi\log q_\phi(\omega|\mathcal{I}^{(k)})].
\end{align*}
Replacing all the expectations in \ding{172}, \ding{173}, and \ding{174}, one obtains the MC estimation in \eqref{eq:nabla-phi}. It should be noted that such MC estimation, though intuitive and straightforward, suffers from high variance \cite{paisley21vbsg}. One effective remedy is the reparameterization technique \cite{Kingma2014}, and the key idea is that one can express the random variable as $\omega=g_\phi(\varepsilon, \mathcal{I})$ (reparameterization), where $\varepsilon$ is an auxiliary variable with independent marginal $p(\varepsilon)$. When generating $\omega^{(k,l)}$, one follows the procedure: $\varepsilon^{(l)}\sim p(\varepsilon)$ and $\omega^{(k,l)}=g_\phi(\varepsilon^{(l)}, \mathcal{I}^{(k)})$. For example, when $\omega\sim \mathcal{N}(\mu, \Sigma^2)$ (univariate Gaussian with mean $\mu$ and variance $\Sigma$), a simple reparameterization is $\omega=\mu+\Sigma \varepsilon$, $\varepsilon\sim \mathcal{N}(0, 1)$. Since this parameterization is beyond the scope of this chapter, we refer the reader to \cite{Kingma2014} for more details on the reparameterization in VB.

\subsubsection{Optimal Access Policy: Approximation and Learning}
\label{subsubsec:policy}
With the trust evaluation process discussed above, we are ready to articulate the access policy $\pi_D$ in ZTD. To simplify our exposition, we take BTE as the underlying trust engine, and our argument also applies to other kinds of trust engines. Recall that the defender's goal is to minimize the objective function $\min_{\pi_D\in \Pi_D}\mathbb{E}\left\{\sum_{t=1}^H \mathbb{E}_{\omega\sim b^t}[u_D(s^t, a_D^t, a_A^t, \omega)]\right\}$. With a slight abuse of notation, let $u_D(s^t, a_D^t, a_A^t, b^t)=\mathbb{E}_{\omega\sim b^t}[u_D(s^t, a_D^t, a_A^t, \omega)]$ be the expected utility under the trust $b^t$. Before articulating how to solve the optimal policy, we first address the solution concept in AIMG, i.e., what is the optimality criterion in this multi-agent decision-making? 

In general, what distinguishes a game problem from a single-agent optimization is that players' optimization problems are entangled. In AIMG, the defender's problem is given by $\min_{\pi_D\in \Pi_D}\mathbb{E}[\sum_{t=1}^H u_D(s^t,a_A^t, a_D^t, b^t)]$, where the attacker's actions $a_A^t$ are involved. To see this more clearly, we expand the expectation expression, and the defender's problem becomes 
\begin{equation}
\label{eq:expand}
    \min_{\pi_D\in \Pi_D}\mathbb{E}_{\pi_D, \pi_A, P, \sigma}\left[\sum_{t=1}^H u_D(s^t,a_A^t, a_D^t, b^t)\right].
\end{equation}
Hence, when the defender determines the access policy, it must take the attacker's move into account and vice versa. From our early argument in BTE, one can view the defender's optimal policy as the minimizer to \eqref{eq:expand} under the anticipated attacker's strategy $\pi_A^*$, i.e., 
\begin{equation}
\label{eq:d-star}
    \pi_D^*\in \argmin \mathbb{E}_{\pi_D, \pi_A^*, P, \sigma}\left[\sum_{t=1}^H u_D(s^t,a_A^t, a_D^t, b^t)\right].
\end{equation}
Then, the remaining question is how to derive such anticipation. From Nash's seminal work \cite{nash1951}, one guiding principle is the unilateral deviation principle, which states that $\pi_A^*$ is a rational anticipation of the attacker's move if the player has no incentive to unilaterally deviate from such strategy, i.e., $\pi_A^*$ solves the minimization problem in \eqref{eq:a-star}. The pair $(\pi_D^*, \pi_A^*)$, given by \eqref{eq:d-star} and \eqref{eq:a-star},constitutes a Nash equilibrium of the AIMG. A formal definition is presented in \Cref{def:pbne}.
\begin{equation}
\label{eq:a-star}
    \pi_A^*\in \argmin \mathbb{E}_{\pi_D^*, \pi_A, P, \sigma}\left[\sum_{t=1}^H u_A(s^t,a_A^t, a_D^t, \omega)\right].
\end{equation}

\begin{svgraybox}
\begin{definition}[Perfect Bayesian Nash Equilibrium]
    \label{def:pbne} Consider the information-asymmetric game with the objectives of the attacker and the defender defined by (\ref{eq:expand}), (\ref{eq:d-star}), and (\ref{eq:a-star}).
    A triple of $\langle \pi_D^*, \pi_A^*, \{b^t\}_{t=1}^H\rangle$ is said to be the perfect Bayesian Nash equilibrium of this game if it satisfies
    \begin{align}
        &\pi_D^*(\cdot|s^t, b^t)\in \argmin \mathbb{E}_{\pi_D, \pi_A^*, P, \sigma}[\sum_{\tau=t}^H u_D(s^\tau,a_A^\tau, a_D^\tau, b^\tau)], \text{ for any } t\in [H], \tag{P1}\label{eq:perfect-1}\\
        &\pi_A^*(\cdot|s^t)\in \argmin \mathbb{E}_{\pi_D^*, \pi_A,  P, \sigma}[\sum_{\tau=t}^H u_A(s^\tau,a_A^\tau, a_D^\tau, \omega)], \text{ for any } t\in [H], \tag{P2}\label{eq:perfect-2}\\
        &  b^{t+1}(\omega)=\left\{\begin{array}{ll}
           \frac{b^t(\omega)\mathbb{P}(\mathcal{I}_D^{t+1}\setminus\mathcal{I}_D^t|\omega)}{\sum_{\omega'\in \Omega} b^t(\omega')\mathbb{P}(\mathcal{I}_D^{t+1}\setminus\mathcal{I}_D^t|\omega')}  & \text{if } \mathcal{I}_{D}^{t+1} \text{is realizable}, \\
            \text{an arbitrary probability distribution}, & \text{otherwise}. 
        \end{array} \right.\tag{C1}\label{eq:consistency}
    \end{align}
    $\mathcal{I}_D^t$ is realizable if there exists $\omega$ such that the conditional probability $\mathbb{P}(\mathcal{I}_D^{t+1}\setminus\mathcal{I}_D^t|\omega)$ is strictly greater than zero. 
\end{definition}
\end{svgraybox}
In \Cref{def:pbne}, \eqref{eq:perfect-1} and \eqref{eq:perfect-2} are refinements of \eqref{eq:d-star} and \eqref{eq:a-star}, respectively. When $t=1$, the refinements coincide with \eqref{eq:d-star} and \eqref{eq:a-star}, leading to a Nash equilibrium. What makes the refinements ``perfect'' is that the $\argmin$ equations hold for any $t\in [H]$. \eqref{eq:perfect-1} and $\eqref{eq:perfect-2}$ are referred to as the perfectness conditions in game theory \cite{fudenberg}, meaning that either player has the incentive to deviate from the equilibrium strategy no matter when (time index $t$) and where (the state $s^t$ and belief $b^t$) they start to play AIMG. Finally, the equilibrium in \Cref{def:pbne} is called Bayesian since the belief is generated in a Bayesian manner. \eqref{eq:consistency} is referred to as the consistency condition: the belief update shall be compatible with the strategy since $\pi_A^*$ is involved in the Bayesian update, see \eqref{eq:bayes-update}. In summary, this perfect Bayesian Nash equilibrium (PBNE) is the solution concept considered in the rest of this chapter, and the optimal access policy refers to the equilibrium strategy $\pi_D^*$ in PBNE.

Solving generic PBNE analytically remains largely an open question, even though recent breakthroughs have shed light on the two-stage Markov game case where the PBNE conditions are rephrased using bilevel-bilinear programming \cite{tao23pot}. The rest of this subsection is devoted to the numerical approximation of PBNE. Similar to solving single-agent Markov decision processes where computational methods can be divided into value-based \cite{tao_multiRL,Tao_blackwell} and policy-based \cite{sutton_PG, tao20causal} approaches, the computation of PBNE (approximately) also follows either value-based, i.e., first approximating the expected utility in \eqref{eq:perfect-1} and \eqref{eq:perfect-2}, or policy-based ones, i.e., searching for the policy directly. The following presents two representative algorithms from the two categories, respectively. 
\runinhead{Belief-Value Iteration}
We begin with the value-based approach. Recall that the perfectness conditions \eqref{eq:perfect-1} and \eqref{eq:perfect-2} are an extension of Bellman's principle of optimality \cite{puterman_mdp} to the multi-agent setting. Naturally, one can transplant the value iteration algorithm \cite{puterman_mdp} in dynamic programming to AIMG. However, value iteration operates using backward induction, whereas the belief update is a forward process (Bayesian update). Consequently, one cannot update the value function (i.e., the expected utility) and the belief simultaneously. 

A variant of value iteration is proposed in \cite{huang2019dynamic} to address the conflict between the value function update and the belief update. The gist is that the updates are performed alternatively: updating the value while fixing the belief and vice versa. We refer to such alternative belief/value updates as belief-value iteration (BVI). Denote by $\mathcal{G}(s, b, u_D,u_A)$ the stage game at the state $s$ under the belief $b$, where the utility functions are $u_D(s,a_A,a_D,b)$ and $u_A(s,a_A,a_D, \omega)$, $\omega\in \Omega$. Let
$\texttt{BayesNash}[\mathcal{G}(s, b, u_D,u_A)]$  be the Bayesian Nash equilibrium operator that takes in the stage game utilities and outputs the equilibrium payoffs $(u_D^*, u_A^*)=\texttt{BayesNash}[\mathcal{G}(s, b, u_D,u_A)]$. The equilibrium payoffs $(u_D^*, u_A^*)$ correspond to the minimum in \eqref{eq:perfect-1} and \eqref{eq:perfect-2}, respectively, with the summations inside the expectations are replaced by the stage game utilities. Mathematically, this equilibrium operator is characterized by bilinear programming \cite{huang2019dynamic,tao23pot}.  

The BVI starts with a belief system initialization $\{b^{(t,0)}\}_{t=1}^H$. For the $k$-th iteration, BVI first fixes the belief system  $\{b^{(t,k)}\}_{t=1}^H$. The $k$-th value iteration is given by the backward induction below. For $t=H,H-1,\ldots, 1$,
\begin{equation}
    \begin{aligned}
    V_D^{(t,k)}(s,b^{(t,k)}), V_A^{(t,k)}(s)&=\texttt{BayesNash}[\mathcal{G}^{(t,k)}(s, b^{(t,k)})], \\
    \mathcal{G}^{(H, k)}(s, b) &= \mathcal{G}(s, b, u_D, u_A),\\
    \mathcal{G}^{(t,k)}(s,b) & = \mathcal{G}(s,b, u_D+V_D^{(t+1,k)}, u_A+V_A^{(t+1,k)}), 
\end{aligned}
\tag{VI}
\label{eq:vi}
\end{equation}
where $\mathcal{G}^{(t,k)}$ is referred to as the subgame starting from time $t$ during the $k$-th iteration, bearing the same spirit of the term ``cost-to-go'' in MDP \cite{puterman_mdp}. The utility function in this subgame is defined in \eqref{eq:subgame}. The attacker's utility $u_A+V_A$ can be defined similarly. We remark that by applying the equilibrium operator \texttt{BayesNash} in \eqref{eq:vi}, the perfectness conditions in \Cref{def:pbne} are satisfied, and $V^{(H,k)}_D$ and $V^{(H,k)}_A$ returned by \eqref{eq:vi} are the equilibrium payoffs of the two players, respectively, under the belief system $\{b^{t,k}\}_{t=1}^H$. 
\begin{equation}
\label{eq:subgame}
    u_D+V_D^{(t+1, k)}(s,a_A,a_D,b^{(t,k)})=u_D(s,a_A,a_D,b^{(t,k)})+\mathbb{E}_{s'\sim P}[V_D^{(t+1, k)}(s', b^{(t+1,k)})].
\end{equation}

Given the value functions, the defender's and the attacker's policies can be determined accordingly by solving $\mathcal{G}^{(t,k)}$, and we denote the resulting policies by $\pi_D^k$ and $\pi_A^k$, respectively. To complete the $k$-th iteration, one needs to update the belief system according to the Bayes rule in \eqref{eq:bayes-update}, which is referred to as belief iteration (BI) in this context shown in \eqref{eq:bi}. This belief iteration guarantees the consistency between the policies $\pi_D^k, \pi_A^k$ and the belief systems $\{ b^{(t,k+1)}\}_{t=1}^H$, as mandated by \eqref{eq:consistency}.  
\begin{equation}
     b^{(t+1,k+1)}(\omega)= \frac{b^{(t,k)}(\omega)\mathbb{P}_{\pi_D,\pi_A}(s^{t+1}|s^t, \omega)}{\sum_{\omega'}b^{(t,k)}(\omega')\mathbb{P}_{\pi_D,\pi_A}(s^{t+1}|s^t,  \omega')}, b^{(1,k+1)}(\omega)=\rho(\omega). \tag{BI}\label{eq:bi}
\end{equation}
This interleaved procedure repeats until no significant improvement is observed in the updated value functions. Even though intuitive, BVI does not offer any convergence guarantees since the operator \texttt{BayesNash} in general is not a contraction mapping \cite{hu03nashQ}. Even assuming it is, we note that the introduction of \eqref{eq:bi} further complicates the analysis, and it remains unclear whether the combination of \eqref{eq:vi} and \eqref{eq:bi} is a contraction mapping. Yet, it is safe to conclude that shall BVI converge, the resulting policies and the belief system must be a PBNE. 
 
\runinhead{Policy Gradient} We now shift the focus from the value-based approach to the policy-based one. For simplicity, we fix the attacker's policy in the sequel and present the policy gradient method \cite{sutton_PG} in reinforcement learning. The key message is that the defender's optimal policy can be learned from sample trajectories using stochastic gradient descent. Consider the defender's problem in \eqref{eq:v-d} where the attacker's strategy is fixed and suppressed. 
\begin{equation}
\label{eq:v-d}
    \min_{\pi_D\in \Pi_D} V_D:= \mathbb{E}_{\pi_D, P, \sigma}\left[\sum_{t=1}^H u_D(s^t, a_A^t, a_D^t, b^t)\right]. 
\end{equation}
Suppose the policy is parameterized by a neural network $\pi_D(\phi), \phi\in \mathbb{R}^n$. Then, one can search for the optimal policy through gradient descent, i.e., $\phi\gets \phi-\nabla V_D(\phi)$ (the learning rate is suppressed). $\nabla V_D(\phi)=\nabla  \mathbb{E}_{\pi_D(\phi), P, \sigma}[\sum_{t=1}^H u_D(s^t, a_A^t, a_D^t, b^t)]$. Recall the MC estimation trick in \eqref{eq:trick}, we rewrite the gradient as in \eqref{eq:pg}, referred to as the policy gradient.  
\begin{equation}
\label{eq:pg}
    \nabla V_D(\phi)=\mathbb{E}_{\pi_D(\phi), P, \sigma}\left[\nabla\log \pi_D(\phi)\sum_{t=1}^H u_D(s^t, a_A^t, a_D^t, b^t)\right].
\end{equation}

Denote a sample trajectory under the policy $\pi_D(\phi)$ (in short, $\phi$) by $\ell(\phi):=\{s^1, a_A^1, a_D^1, u_A^1, u_D^1, o^1, \ldots, s^H, a_A^H, a_D^H, u_A^H, u_D^H, o^H\}$, where $u_D^t=u_D(s^t, a_A^t, a_D^t, b^t)$, $b^t$ is derived using the Bayes rule in \eqref{eq:bayes-update}. Then, an unbiased estimate of $\nabla V_D(\phi)$, denoted by $\widehat{\nabla} V(\phi)$ is constructed as $\widehat{\nabla} V(\phi)=\nabla\log \pi_D(\phi)\sum_{t=1}^H u_D^t$. Denote by $u_D(\ell)=\sum_{t=1}^H u_D^t$ the empirical return of the sample trajectory. One common practice to reduce the variance of the MC estimate $\widehat{\nabla} V(\phi)$ is to collect a batch of trajectories $\{\ell^{(k)}\}_{k=1}^K$ and take the average: $\widehat{\nabla} V(\phi)=1/K \sum_{k=1}^K\nabla\log \pi_D(\phi)u_D(\ell^{(k)})$. Starting from an initialization $\phi^0$, one need first implement the policy $\pi_D(\phi^0)$ in a simulated network system \cite{kim23dt} and collect a batch of trajectories $\{\ell^{(k)}\}_{k=1}^K$. Then, the policy is updated using the policy gradient discussed above. The procedure repeats until the parameter $\phi^k$ stabilizes. Since policy gradient is a first-order method, it is only guaranteed to converge to the first-order stationary point where $\nabla V_D(\phi)=0$. Even though this first-order point may not be the exact equilibrium point, it often leads to satisfying defense policy, as observed in the literature \cite{tao22sampling}. 

\subsection{Generalizability, Explainability, and Accountability of Learning-based Zero-Trust Defense}
\subsubsection{Reinforcement Learning and Explainable Defense}
\label{subsec:xrl}
Even though RL leads to a theoretically guaranteed approach to learning the ZTD policy,  the missing part is that the learned policy, i.e., the model weights of the neural network, remains a black box and is difficult for human operators to comprehend. The explainability of RL (XRL), as an emerging field devoted to casting light on the inner workings of RL agents, has gained momentum across various research communities.  Since XRL is still in its infancy, there is no consensus over the exact definitions of explainability, and most of the current endeavors try to explain the actions of RL agents \cite{dazeley23xrl}. Following this line of research, we discuss the explainability of the optimal access policy learned by RL in the following, which addresses the question: 
\begin{center}
    \textit{How does the RL policy grant or deny access based on the trust evaluation?} 
\end{center}
Our XRL approach exploits the mathematical structure of the AIMG and utilizes non-parametric policy learning, i.e., the RL policy is expressed in closed form without involving neural networks \cite{ge22threshold, tao23ztd}. Hence, our XRL study is more aligned with the interpretability of the RL policy, indicating that the intrinsic logic of the defense mechanism is transparent and easy to understand rather than a post-hoc property.       

The gist of the explainability in ZTD is that the optimal policy is of a threshold form \cite{ge22threshold}. Consider the lateral movement case in \Cref{subsec:lateral} as an example, where the type space and the defense action space are binary: $\Omega=\{0, 1\}$ (0-legitimate user, 1-attacker) and $\mathcal{A}_D=\{0, 1\}$ (0-active defense, 1- inactive). In this example, the belief $b$ resides in the two-dimensional probability simplex, which can be uniquely determined by its entry $b(0)$. We refer to $b(0)\in[0,1]$ as the trust score, implying the likelihood of the user is legitimate. A threshold policy $\pi_D(b)$ is defined in \eqref{eq:threshold}, and the threshold is given by $\tau$. As its name suggests, the defense remains idle as long as the trust score is above the threshold, while it is activated once the trust score is below the critical value. 
\begin{equation}
\label{eq:threshold}
    \pi_D(b)=\left\{\begin{array}{ll}
       0,  & 0\leq b(0) \leq \tau, \\
       1,  & \tau < b(0) \leq 1.
    \end{array}\right.
\end{equation} 
% \begin{figure}[!ht]
%     \centering
%     \includegraphics[width=0.8\textwidth]{figs/policy.pdf}
%     \caption{An illustration of the threshold policy. Once the trust score falls below the threshold 0.5, the access is denied, and the defense is activated.}
%     \label{fig:threshold}
% \end{figure}

The advantage of this threshold policy is self-evident: it is a white box clearly displaying how the trust evaluation is utilized. The same policy gradient method presented above also applies to the learning of thresholds. Even though the gradient $\nabla_\tau \pi_D$ does not acquire a closed form, one can leverage the simultaneous perturbation stochastic approximation (SPSA) to estimate the gradient \cite{ge22threshold,tao23ztd}. The threshold form in \eqref{eq:threshold} also extends to the finite-action case, where $|\mathcal{A}_D|-1$ threshold values partition the interval $[0,1]$ into $|\mathcal{A}_D|$ subintervals (the type space is still binary).

\subsubsection{Meta-Learning and Generalizable Defense }
The limitation of the threshold policies is concerned with generalization ability. The optimal policy (or equivalently, threshold) trained in one network setup cannot deal with another scenario where the system vulnerabilities are different from the training setup. To facilitate our discussion, denote by $\theta\in \Theta$ the network system configuration that can affect the system transition $P$ (or the observation function $\sigma$)  under this configuration. Using the notations in \Cref{def:mg}, the defender now faces a family of games, and the transition function $P_\theta$ of each game is parameterized by $\theta$ subject to a distribution $p(\theta)$. We refer to each game under parameter $\theta$ as an attack scenario. The policy trained for the scenario $\theta$ does not generalize well to $\theta'$, leading to ineffective ZTD. 

To equip ZTD with generalizability under information asymmetry, a scenario-agnostic ZTD (SA-ZTD) is proposed in \cite{tao23ztd}, creating a generalizable ZTD capable of handling new attack scenarios unseen in the training phase. SA-ZTD rests on meta-learning, an emerging learning paradigm that aims to learn a learning strategy using training data \cite{meta_survey}. In the face of a new scenario unseen in the training phase, the obtained learning strategy enables the defender to learn a new defense on the fly using far fewer data than from scratch. This idea of defending on the fly is also explored in adversarial machine learning leading to impressive defense performance \cite{li23metasg}. Since real-world applications involve a large (possibly infinite) number of attack scenarios, it is intractable to learn the optimal policy for each scenario. Powered by meta-learning, SA-ZTD uses only a handful of known scenarios, more precisely, sample trajectories from these scenarios.  Hence, the word ``agnostic,'' whose root means ``not known,'' is used to emphasize that the adaptation ability is acquired without knowledge of the network configuration of every scenario.

Two pillars of SA-ZTD are the meta policy $\pi_{meta}$ and the adaptation mapping $\Psi: \Pi_D\times \Theta\rightarrow\Pi_D$. The adaptation mapping corresponds to the learning strategy mentioned earlier that adapts the meta policy to a new defense $\Psi(\pi_{meta}, \theta)$ when facing a new scenario $\theta$. A formal definition of SA-ZTD is given in \cite{tao23ztd}, which we restate in \Cref{def:sa-ztd}. 
\begin{svgraybox}
\begin{definition}[SA-ZTD]
    \label{def:sa-ztd}
    A pair $\left\langle \pi_{meta}, \Psi \right\rangle$ is said to be a scenario-agnostic zero-trust defense (SA-ZTD) with respect to a scenario distribution $p\in \Delta(\Theta)$ if the pair solves for the minimization problem 
   \begin{align}
       \min_{\pi,\Psi}\mathbb{E}_{\theta\sim p}[V_D(\Psi(\pi,\theta))].
       \label{eq:meta-obj}
   \end{align}
\end{definition}
\end{svgraybox}
Similar to empirical risk minimization (ERM) \cite{vapnik1999nature, liu2020communication}, a solution to \eqref{eq:meta-obj} is obtained by solving the sample average approximation:
\begin{align}
    (\pi_{meta},\Psi)\in \argmin \frac{1}{|\widehat{\Theta}|}\sum_{\theta\in \widehat{\Theta}}V_D(\Psi(\pi,\theta)),
    \label{eq:meta-emp}
\end{align}
where $\widehat{\Theta}\subset\Theta$ is a finite collection of scenarios i.i.d. sampled from $p\in \Delta(\Theta)$. The term ``agnostic'' points to the fact that the exact scenario distribution $p$ is usually unknown in security practice and often replaced by an empirical distribution provided by security datasets, such as the data from MITRE ATT\&CK \cite{strom2018mitre} considered in \cite{tao23ztd}. In summary, the training of SA-ZTD does not explicitly require the domain knowledge of each attack scenario, such as the system configuration and the observation functions. 

Since the function class $\{\Psi|\Psi:\Pi_D\times \Theta\rightarrow\Pi_D\}$ is infinite-dimensional, directly seeking an adaptation mapping through \eqref{eq:meta-obj} [or \eqref{eq:meta-emp}] is intractable. One remedy is to restrict the focus to the parameterization class where the mapping is parameterized by  $\gamma\in \mathbb{R}^n$, $n\in \mathbb{Z}_{+}$. For example, $\Psi_\gamma$ can be parameterized by recurrent neural networks, where $\gamma$ is the model weights and the optimal adaptation is determined by training algorithms \cite{hochreiter01meta-recurrent}.  Another well-accepted parameterization is the gradient-based adaptation: $\Psi_\gamma(\pi, \theta):=\pi-\gamma \nabla V_D(\pi)$, and $\gamma$ is the gradient step size to be optimized \cite{meta-sgd}. 

To arrive at an explainable SA-ZTD, one can pick the gradient-based adaptation, as it naturally applies to the non-parametric threshold policies discussed in \Cref{subsec:xrl}.  To be consistent with previous notations, we replace $\pi$ with $\tau$ whenever speaking of threshold policies, where the $\tau$ denotes the threshold value. The minimization problem in \eqref{eq:meta-obj} turns into 
\begin{equation}
    \label{eq:meta-tau}
    \min_{\tau\in [0,1]}\mathbb{E}_{\theta\sim p}[V_D(\operatorname{Proj}_{[0,1]}\{\tau-\gamma\nabla V_D\})].
\end{equation}
The resulting meta policy, as the minimizer to \eqref{eq:meta-tau}, takes the threshold form that is explainable to human operators, increasing the accessibility and transparency of learning-based ZTD. As argued in \cite{tao23ztd}, the policy gradient method is still applicable to \eqref{eq:meta-tau}. Even though the computation expenditure in SA-ZTD is higher than the vanilla RL policy in \eqref{eq:v-d}, the meta policy can adapt to a variety of new scenarios without training from scratch. 

% We finally conclude this subsection 
% \begin{figure}[!ht]
%     \centering
%     \begin{subfigure}{0.49\textwidth}
%         \includegraphics[width=\textwidth]{figs/pup_pi_difference.eps}
%     \caption{Meta policy under different system vulnerability distributions.}
%     \label{fig:enter-label}
%     \end{subfigure}
%     \hfill
%     \begin{subfigure}{0.49\textwidth}
%         \includegraphics[width=\textwidth]{figs/pup_adapt.eps}
%     \caption{Adapted policy $\tau_{adapt}$ after one-step gradient update.}
%     \label{fig:enter-label}
%     \end{subfigure}
%     \hfill
%     \begin{subfigure}{0.49\textwidth}
%         \includegraphics[width=\textwidth]{figs/pap_pi_difference.eps}
%     \caption{Meta policy under different attacker stealthiness distributions.}
%     \label{fig:enter-label}
%     \end{subfigure}
%     \hfill
%     \begin{subfigure}{0.49\textwidth}
%         \includegraphics[width=\textwidth]{figs/pap_adapt.eps}
%     \caption{Adapted policy $\tau_{adapt}$ after one-step gradient update.}
%     \label{fig:enter-label}
%     \end{subfigure}
% \end{figure}

\subsubsection{Accountability}
The accountability of machine-learning-based ZTD (ML-ZTD) refers to the responsibility and answerability of those involved in the design, development, deployment, and use of machine learning or artificial intelligence technologies in general. Accountability aims to ensure that ML-ZTD is developed and utilized in a manner that is ethical, transparent, and fair. What distinguishes accountability of ZTD in 5G networks from other AI systems is the focus on accountability in system engineering, which encompasses three key aspects: responsibility, detectability, and attribution. 
\runinhead{Responsibility} Accountability rests on the acknowledgment that individuals and organizations involved in ML-ZTD development and deployment have responsibility for the ZTD's behavior and impact on the network system. Specifically, this responsibility revolves around the question of whether each component involved in ZTD architecture, such as the security machinery, the trust engine, and the access policy, contributes to an ethical, transparent, and fair operation in the network. To be more precise, this responsibility provides compliance requirements and failure standards for each component. 
\runinhead{Detectability} Responsibility gives the rule book, and the next question to address is whether ZTD operation violates the compliance requirements. Mathematically, the detectability question pertains to statistical inference, such as hypothesis testing and VB methods, where one infers the ground truth (violation) from collected data. Yet, ZTD in 5G networks is a game problem, see \Cref{def:mg}, where the strategic decision-maker can evade the detection, which must be taken into account when inspecting the ZTD operation. Game theory naturally provides a system-science viewpoint on the detectability question in multi-agent systems, where the incentives, capabilities, and private information of the investigator and the investigatee can be captured through the AIMG in \Cref{def:mg}.  This game-theoretic viewpoint leads to a strategic detection framework.   

\runinhead{Attribution} No node is an island in large-scale complex 5G networks, and one failing node or component may spur a chain reaction over the network and the ZTD system. When facing a cascading failure in the network defense, one needs to identify the root cause and upgrade the ZTD accordingly. One shall not confuse detection with attribution, even though both of them aim to identify the malfunctioning part of the ZTD and the network system. However, detection addresses the question ``where it is'', whereas attribution focuses on ``why it is such.'' Mathematically, attribution amounts to a causal inference task \cite{tao20causal}, where the casual relationship among  random variables is established using data.   
}

{
\section{Decision-dominance Defense}
\label{sec:dd}
% why we need decision dominance defense 

While ZTD provides us with a comprehensive framework for trust evaluation and access policy, the networked entities still face multi-stage persistent cyber threats. 
Therefore, it is crucial to adopt an integrated defense approach that recognizes the intrinsic value of the cyber defense chain and the fundamental principles of zero trust. 
Decision dominance defense (D$^3$), which conceptualizes the interactions of cyber defense/kill chain as a stochastic process, forms the backbone of the holistic defense mechanism, with zero trust defense acting as a critical component at every stage. 
By treating the cyber defense chain as a dynamic system, we acknowledge the unpredictable nature of cyber threats and the need for proactive decision-making based on real-time information. 
By incorporating zero trust principles throughout this process, from initial access controls to ongoing monitoring and incident response, we create a robust and resilient defense model that embraces uncertainty, eliminates blind spots, and ensures continuous protection against the relentless onslaught of cyber threats.
% what is decision dominance defense (introductory)

Understanding the intricacies of an attack is crucial for developing effective defense strategies. 
A traditional Lockheed Martin Kill Chain \cite{yadav2015technical,khan2018cognitive} usually outlines seven distinct stages that malicious actors typically follow. These stages include Reconnaissance, where attackers gather information on potential targets; Weaponization, where they create malicious tools or payloads; Delivery, the method through which the attack is transmitted; Exploit, where vulnerabilities are leveraged to gain access; Installation, the establishment of a foothold within the target system; Command \& Control, the creation of communication channels for remote control; and finally, Actions on Objectives, where the attacker achieves their intended goals within the compromised system. 
Comprehensively analyzing and understanding each stage of the Kill Chain requires the defender to effectively engage with adversaries while minimizing the time it takes for an attack to unfold. A proactive cyber defense chain (e.g., \cite{huang2019adaptive,heckman2015denial}) aims to disrupt and curtail the attacker's progress at each stage of the Kill Chain, reducing their opportunity to inflict significant damage. D$^3$ integrates real-time threat intelligence, advanced analytics, and rapid response mechanisms, including monitoring, detection, response, and attribution, maximizing the abilities to mitigate and neutralize the threats, actively impeding the attacker's progress and shortening the overall time it takes for an attack to materialize.
It empowers the 5G network defender to take a more active role in their defense, enabling them to stay one step ahead of the adversary and significantly enhance their resilience against evolving cyber threats.

% what is that the decision dominance does 

The essence of D$^3$ is the critical timing of cutting off the cyber kill/defense chain. In MWD scenarios, while the general concept of understanding, deciding, acting, and assessing fast still holds (i.e., strangling the threats in its cradle), one must take the real-time warfare conditions and game-theoretic thinking into consideration, ``knowing oneself and knowing the enemy''. 
Therefore, in the sequel, we formalize D$^3$ as a Dynkin's type of optimal stopping game acting on a Markov chain of multi-stage cyber-attacks/defense \cite{gore2017markov}, and characterize the equilibrium strategy between the two competitive parties. While our model is built upon ZTD components, the notations should not be confused with the previous section.

\subsection{D$^3$ as Dynkin's Game}

By convention, let $(\Omega, \mathcal{F}, \mathbb{P})$ be the probability space. Denote the time index during a lifecycle of the interactions between the cyber kill/defense chain by $t = 1, \ldots, T$. 
Let $(X_t)_{0 \leq t \leq T}$ be a Markov process modeling the cyber threats, living in space $(\mathcal{X}, \mathcal{G})$, and are adapted to the filtration $\mathbb{F} = (\F_t)_{0 \leq t \leq T}$ with transition kernel $\Pk$. The Markovian state captures the identifiable elements in the system, e.g., it can represent the Structured Threat Information eXpression language (STIX) that facilitates this effort \cite{gore2017markov}. The collection of STIX-type data requires active interactions between the two parties.

% Let $\mathcal{E}(\mathcal{X})$ be the set of all bounded $\mathcal{B}(\mathbb{R})/\mathcal{G}$-measurable functions on $(\mathcal{X}, \mathcal{G})$.
We are given three payoff functions $\phi, \zeta, \psi: \mathcal{X} \to \mathbb{R}$ that capture the cyber risk given system states, where from the defender's perspective, (the attacker's perspective would be the opposite,)
\begin{itemize}
    \item[1.] $\phi$ is the early termination payoff,  which is activated when the cyber defender actively terminates the persistent monitoring/detection and resets the system credential before the malicious operations, including data exfiltration, denial of service, and delivery of ransomware, etc. are executed;
    \item[2.] $\psi$ is the late response payoff, which is activated when the cyber defender responds to the data exploitation and command \& control actions without summarizing the monitoring/detection phase. 
    \item[3.] $\zeta$ is the confrontation payoff, which is activated when both parties have extracted information through lateral movement/monitoring and engaging, etc., and perform attack/defense actions at the same stages. 
\end{itemize}
It is reasonable to assume that $ \min (\psi, \phi) \leq \zeta \leq \max(\psi, \phi)$, since the confrontation often happens when attackers and defenders both 
have neutralized assessments for the system, it sits in between the worst and best payoffs.

Here, for simplicity, we first consider the case where the information is symmetrical between the network operator/defender and the attacker, i.e., both parties have access to the state and utility information. However, this formalism shall not exclude the cases where the information is asymmetric and/or the utility functions are unknown/uncertain to one of the parties.

On top of the lower-level cyber threats/defense operations, we define stopping times
$\tau, \sigma: \Omega \to \{ 0, \ldots, T \}$ to capture the termination decisions for both parties. 
With the assumption that both the attacker and the defender have access to the system state $X_t$, $\tau, \sigma$ are $\mathbb{F}$-measurable.
Denote the set of $\mathbb{F}$-stopping times by $\mathcal{T}: = \{ 0 \leq \tau \leq T:  \{\tau(\omega) \leq k\} \in \mathcal{F}_k  \ \forall k \in [T], \forall \omega \in \Omega  \}$. 
Moreover,  we expect there to be a $2^{[T]}/\mathcal{G}$-measurable map $\tau: \mathcal{X} \to [T]$, where $[T] = \{0,\ldots, T\}$, such that the defender/attacker will make termination decisions based on the information extracted from $X_t$, without awareness of each other's stopping decisions.

For stopping times $\tau, \sigma$, the value/cost function for the defender/attacker is defined as: 
\begin{equation} \label{eq:payoff}
    V^{\tau, \sigma}(x) = \mathbb{E}_{x} [H(\tau, \sigma)] =  \mathbb{E}_{x}\left[\phi(X_\tau)\mathds{1}_{\{ \tau < \sigma\}} + \psi(X_\sigma)\mathds{1}_{\{ \tau > \sigma\}} + \zeta(X_\tau) \mathds{1}_{\{ \tau = \sigma\}}\right],
\end{equation}
where $H(\tau, \sigma): \T \times \T \times \Omega \to \mathbb{R}$ is the random payoff of stopping strategies $\tau$ and $\sigma$, $\mathbb{E}_x$ is the conditional expectation operator with respect to the transition kernel $\Pk_x$, i.e., there is an operator $\mathscr{T}$ that is a semi-group, such that for any $\mathcal{B}(\mathbb{R})/\mathcal{G}$-measurable function $g$ and $t= 0, \ldots, T$, $$ \mathscr{T}^t g (x ) := \mathbb{E}_x [ g ( X_t)] = \underbrace{\int_\mathcal{X} \ldots \int_\mathcal{X}}_{t \text{ times}} g(x_t)  d \Pk_{x_{t-1}}(x_t) \ldots d \Pk_x ( x_1). $$ 
In practice, the convolutional integral is hard to compute directly. Instead, we can leverage sampling methods such as Markov Chain Monte-Carlo (MCMC) to approximate the conditional expectation. 

\Cref{def:ddg} summarizes our game-theoretic formalism.

\begin{svgraybox}
     \begin{definition}[Decision Dominance Game] \label{def:ddg}
          A tuple $( \mathcal{X}, \Pk, \phi, \zeta, \psi, \mathcal{T})$ encapsulates a Decision Dominance Game (DDG) if it satisfies the following:
          \begin{itemize} 
              \item there exists a Markov process $(X_t)_{0 \leq t \leq T}$ that lives in $(\mathcal{X}, \mathcal{G})$ with transition kernel $\Pk$, which can be extracted as cyber threats information;
              \item $\phi, \zeta,$ and $\psi$ are payoff functions mapping from $X_t$ to $\mathbb{R}$, $ \phi, \zeta, \psi  \in \mathcal{E}(\mathcal{X})$, which is the set of all bounded $\mathcal{B}(\mathbb{R})/\mathcal{G}$-measurable functions on $(\mathcal{X}, \mathcal{G})$. Further, $\min(\phi, \psi) \leq \zeta \leq \max(\phi, \psi)$ on $\mathcal{X}$;
              \item at each stage $t$, both parties pick a stopping strategy from space $\T_t := \{ t \leq \tau \leq T:  \{\tau(\omega) \leq k\} \in \mathcal{F}_k  \  \forall k \in [T], \forall \omega \in \Omega  \}$ to decide whether to stop or continue the kill/defense chain.
              \item at each stage the utility function of the defender is 
              \begin{equation*}
                   H(\tau_t, \sigma_t) = \phi(X_{\tau_t}) \mathds{1}_{\{ \tau_t < \sigma_t\}} + \zeta(X_{\tau_t}) \mathds{1}_{\{ \tau_t = \sigma_t\}} +  \psi(X_{ \sigma_t })\mathds{1}_{\{ \tau_t >  \sigma_t\}}, 
              \end{equation*}
              while the attacker attains $-H(\tau_t, \sigma_t)$.
          \end{itemize}
     \end{definition}
\end{svgraybox}
\begin{figure}
    \centering
    \includegraphics[width = \textwidth]{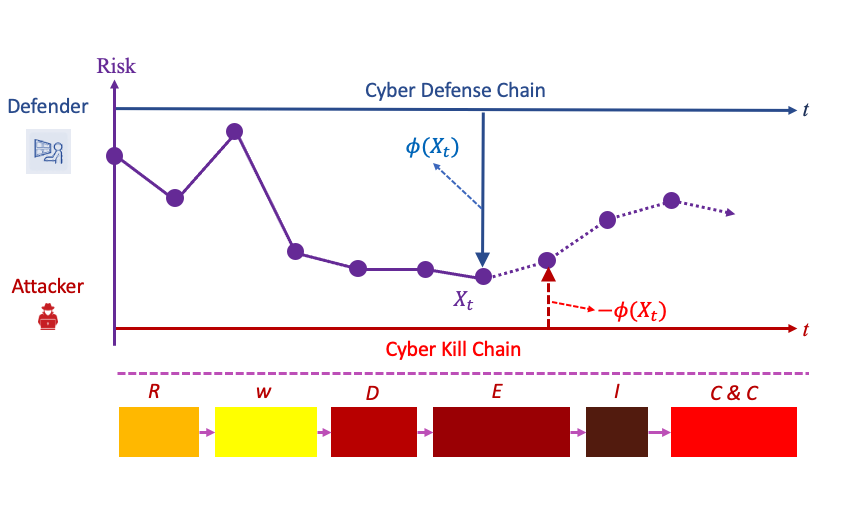}
    \caption{An illustration of the cyber kill/defense chain interaction. In this case, at time $t$, the system state has evolved into $X_t$, the defender cuts off the chain interaction earlier than the attacker and gets payoff $\phi(X_t)$, while the attacker gets $-\phi(X_t)$ since she plans to stop at the next time step}
    \label{fig:killchain}
\end{figure}
\Cref{fig:killchain} gives an example of the DDG outcome. 
The solution concept of a DDG is given in \Cref{def:ne}.
\begin{svgraybox}
    \begin{definition}[Decision-Dominance Equilibrium (DDE)] \label{def:ne}
    A pair of stopping time strategies $(\tau^*, \sigma^*) \in \mathcal{T} \times \mathcal{T} $ is a Decision-Dominance Equilibrium (DDE) if for all initial state $x \in \mathcal{X}$, it satisfies the minimax condition:
    \begin{equation} \label{eq:dde}
\begin{aligned}
    V^{\tau^*, \sigma^*} (x) & = \essp_{\sigma \in \T}  \essf_{\tau \in \T} V^{\tau, \sigma}(x) \\ 
       & = \essf_{\tau \in \T} \essp_{\sigma \in \T}  V^{\tau, \sigma}(x).
\end{aligned}
\end{equation} 
\end{definition}
\end{svgraybox}

The existence of such a value function, however, is a non-trivial question, as we are looking for a pure strategy Nash equilibria in an infinite-dimensional space ($\T \times \T$), Von-Neumann's Minimax theorem does not apply here. However, under certain conditions, we are able to show that a DDG with information symmetry always admits a value function, which is unique up to a state-wise constant translation.

We know from Dynkin's result \cite{kingman1976discrete} that when $\phi \leq \zeta \leq \psi$ on $x \in \mathcal{X}$, there exists a value process 
\begin{equation} \label{eq:standardres}
\begin{aligned}
    V_t & = \min\{ \psi(X_t), \max\{ \phi(X_t), \mathbb{E}[ V_{t+1} | \F_t ]\}\} \\
    & = \max\{ \phi(X_t), \min\{ \psi(X_t), \mathbb{E}[ V_{t+1} | \F_t ]\}\},
\end{aligned}
\end{equation}
and the equilibrium strategies capture $V_t$'s hitting times of the upper/lower limits. 
However, the ordered-payoff assumption is hard to verify in the context of MDW, a more reasonable assumption, as has been discussed before, is $\min(\phi, \psi) \leq \zeta \leq \max(\phi, \psi)$ on $\mathcal{X}$. In addition, the ubiquitous information asymmetry in cyberspace oftentimes makes the derived equilibrium strategies inapplicable. 

Therefore, in the sequel, we dive into the more general case defined as in \Cref{def:ddg}, and lay out some essential analytical characterization for the equilibrium value process; further, we give a rough description for the case under information asymmetry.

\subsection{Equilibrium Strategies for D$^3$}

In this section, we investigate the existence and characterization of the DDE in two different cases under a symmetric information structure and then discuss an extension. 
The first case is when the early termination payoff $\phi$ dominants the late termination payoff $\psi$,  which we call \textit{adversarial dominance}, as in this case, the outcome of engaging in the long term favors the adversary.
The second case is called \textit{defense dominance}, where the late termination payoff $\psi$ dominates the early termination payoff $\phi$. Hence, the defender is able to endure the kill/defense chain interactions longer than the adversary does.
\subsubsection{Case I: Adversarial Dominance}

%\textcolor{red}{why is it a definition?}
 
    Under the Adversarial Dominance Condition (ADC), the payoff functions satisfy the ordered condition $\psi \leq \zeta \leq \phi$ for all system states $x \in \mathcal{X}$. In this case, at any state $x \in \mathcal{X}$, the defender aims to investigate the kill chain for a proper period of time while trying to terminate the operations faster than the attacker, as it is more costly to wait for the attacker to exploit the vulnerabilities by doing Command \& Control than to shut down the service and reset the credentials. This is also called \textit{first-mover advantage}, that is, the defender has the incentive to end the game faster than the opponent.

We shall proceed with the analysis by giving a constructive sequence of equilibrium values. 
To this end, we investigate the $t\in [T]$ stage problem through backward induction and let $\{V^t_n\}_{n=0}^t$ be the equilibrium processes attained by stopping at no more stage $t$. 
At $t \in [T]$, both parties have to choose confrontation, thus at the final stage, the payoff is $\zeta(X_t)$; at $n \in [t-1]$, either they both stop and get payoff value $\zeta(X_n)$, or wait for the next round, in which case the defender has to judge if the termination values $\phi(X_n)$ is higher than the expected engaging values $\mathbb{E}\left[V_{n+1}^{t} | \F_{n}\right]$, given that the attacker chooses to engage. Mathematically, we have the value processes for arbitrary $t \in [T]$,
\begin{equation}\label{eq:constructv}
    \begin{aligned}
         V^t_t & = \zeta(X_t), \\
         V^t_n & = \operatorname{val}\left[\begin{array}{cc}\zeta(X_n) & \phi(X_n) \\ \psi(X_n) & \mathbb{E}\left[V_{n+1}^{t} | \F_{n}\right]\end{array}\right] ,\quad\quad \text{ for } n = t-1, \ldots, 0.
    \end{aligned}
\end{equation}
where $\operatorname{val}(\cdot)$ stands for a special value operator of the matrix game, which we interpret as: 
\begin{align*}
    V^t_n =\begin{cases} \mathbb{E}\left[V_{n+1}^{t} | \F_{n}\right] & \text { if }\phi(X_n) < \mathbb{E}\left[V_{n+1}^{t} | \F_{n}\right], \\ \zeta(X_n) & \text { otherwise }.\end{cases}
\end{align*}

It turns out that the value processes possess the monotone property (Lemma \ref{lemma:monotone}).

   \begin{lemma}
    \label{lemma:monotone}
     For every $n, t \in [T]$ such that $n \leq t$, one has that the equilibrium value processes defined as in \eqref{eq:constructv} satisfy
     \begin{equation*}
         V^n_n \leq V^t_n \leq V^{t+1}_n  .
     \end{equation*}
\end{lemma}

One can show \Cref{lemma:monotone} with an induction argument. Let the event $E_n: = \{ \omega:  \phi(X_n) < \mathbb{E}[\zeta( X_{n+1}) | \F_n]\}$ be when the next round expected confrontational payoff is higher than the current early termination payoff. 
Consider the base case; it follows that at any stage $t \in [T-1]$, since the next round both parties need to terminate, it is reasonable for the defender to choose to terminate if the early termination payoff is higher than the expected confrontational payoff. Thus,
 \begin{align*}
     V^{t+1}_t & = \begin{cases}
        \mathbb{E}[ \zeta(X_{t+1})| \F_t]  \quad &\text{ on } E_t, \\
        \zeta(X_t) \quad & \text{ on } E_t^c ,
     \end{cases}  \\
      & \geq \begin{cases}
            \phi (X_t)  \quad & \text{ on }  E_t, \\
        \zeta(X_t) \quad & \text{ on } E_t^c ,
      \end{cases} \\
      & \geq \zeta(X_t)   = V^t_t. 
 \end{align*}

Now we assume that $V^{j+k-1}_j \leq V^{j+k}_j$ for some arbitrary stage $1 \leq k \leq T-1$ and for all $j \in [T-k]$, then, for $t \in [T-k-1]$, 
\begin{equation*}
\begin{aligned}
    V^{t+k}_t & = \operatorname{val}\left[\begin{array}{cc}\zeta(X_t) & \phi(X_t) \\ \psi(X_t) & \mathbb{E}\left[V_{t+1}^{t+k} | \F_t\right]\end{array}\right] \\
    & \leq \operatorname{val}\left[\begin{array}{cc}\zeta(X_t) & \phi(X_t) \\ \psi(X_t) & \mathbb{E}\left[V_{t+1}^{t+k+1} | \F_t\right]\end{array} \right] \\
    & =  V^{t+k+1}_t   .
\end{aligned}
\end{equation*}
Hence, the monotonicity follows by the induction argument.

That  $V_k^t$ being increasing in $t$ gives off two signals; the first is that due to the Monotone Convergence theorem for $\mathbb{E}[\cdot | \F_t ]$, there exists a limit for $V_k^t$ if we consider the infinite-stage problem ($t \to \infty$); the second is that the dominating strategy can be obtained when the stopping stage is not constrained, up to time $T$.

Now we define two stopping times, for $t \in [T]$, 
\begin{align*}
    \bar{\tau}_t  & = \inf \{ t \leq k \leq T | V^T_k = \zeta(X_k)\} , \\
    \bar{\sigma}_t & = \inf \{ t \leq k \leq T | V^T_k = \zeta(X_k)\}. 
\end{align*}
The significance of $(\bar{\tau}_t, \bar{\sigma}_t )$ is given in \Cref{thm:eqst}.
 %\Cref{thm:eqst} states that in the adversarial dominance environment, $(\bar{\tau}_0, \bar{\sigma}_0)$ are the equilibrium strategies.
 
\begin{theorem} \label{thm:eqst}
Under ADC, the following statements hold for arbitrary initial state $x \in \mathcal{X}$:
\begin{itemize}
    \item[i)] For every $t \in [T]$, and all $\tau \in \T_t, \sigma \in \T_t$,
    \begin{align*}
        \mathbb{E}[ H(\tau, \bar{\sigma}_t) |\F_t] \leq  V^T_t =  \mathbb{E}[V^T_{\bar{\tau}_t \wedge \bar{\sigma}_t}|\F_t] =  \mathbb{E}[  H( \bar{\tau}_t, \bar{\sigma}_t) | \F_t]  \leq  \mathbb{E}[ H(\bar{\tau}_t, \sigma) |\F_t] .
    \end{align*}
    \item[ii)] At every time $t \in [T]$, a pair $(\bar{\tau}_t, \bar{\sigma}_t)$ is an equilibrium point for that time step $t$, and a DDE value corresponding to $(\bar{\tau}_0, \bar{\sigma}_0)$ is given as
    \begin{align*}
        \mathbb{E}[V^T_0] = \mathbb{E}[V^T_{\bar{\tau}_0 \wedge \bar{\sigma}_0 }]  = \mathbb{E}[ H (\bar{\tau}_0 , \bar{\sigma}_0)] .
    \end{align*}
\end{itemize}
    
\end{theorem}

\begin{proof}
    Fix a $t \in [T]$ arbitrarily. We have that, if $k \in \{ t, \ldots, \bar{\sigma}_t\}$, by definition of $\bar{\sigma}_t$, we have
    \begin{equation*}
        V^T_k = \mathbb{E}[V^T_{k+1} | \F_k ].
    \end{equation*}
    Thus, the sequence $\{ V^T_{k \wedge \bar{\sigma}_t}, k \geq t\}$ is a regular Martingale, so that $V^T_t = \mathbb{E}[ V^T_{\tau \wedge \bar{\sigma}_t }|\F_t]$ for any $\tau \in \T_t$, by Doob's optional sampling theorem. 
    Since $V^T_{\bar{\sigma}_t} = \zeta(X_{\bar{\sigma}_t}) \geq \psi(X_{\bar{\sigma}_t})$, if $\bar{\sigma}_t \leq \infty$ and $V^T_k \geq \phi(X_k)$ if $\bar{\sigma}_t > k$, it follows that:
    \begin{align*}
        V^T_t &= \mathbb{E}[ V^T_{\tau \wedge \bar{\sigma}_t} | \F_t] \\
         & = \mathbb{E}[ V^T_{\tau} \mathds{1}_{\{\tau < \bar{\sigma}_t\}} +  V^T_{\bar{\sigma}_t} \mathds{1}_{\{ \bar{\sigma}_t \leq \tau\}}   | \F_t] \\
        & \geq \mathbb{E}[ \phi(X_\tau) \mathds{1}_{\{\tau < \bar{\sigma}_t\}} +  \psi(X_{\bar{\sigma}_t}) \mathds{1}_{\{ \bar{\sigma}_t < \tau\}} +\zeta(X_{\bar{\sigma}_t}) \mathds{1}_{\{ \bar{\sigma}_t = \tau\}} |\F_t ] \\
        & = \mathbb{E} [ H(\tau, \bar{\sigma}_t) | \F_t] .
    \end{align*}
     A symmetric argument can be applied to prove the $\leq$ side for all $\sigma \in \T_t$. By letting $t = 0$ we arrive at the conclusion.
\end{proof}

\Cref{thm:eqst} i) implies that for every subgame starting from time $t$, the equilibrium strategy is always a threshold strategy for both parties, where the threshold needed to be computed is $\mathbb{E}[V^T_{t+1} | \F_t]$. Both parties have incentives to stop only when $\phi(X_t)$ is hitting the threshold. 
ii) states that in the adversarial dominance environment, $(\bar{\tau}_0, \bar{\sigma}_0)$ are the equilibrium strategies. However, the determination of the 
equilibrium value sequence $V^T_t$ is computationally intractable, as one would have to construct the random variables backwardly according to \eqref{eq:constructv}, enumerating over the filtration sets.

Therefore, it is crucial to generalize the above arguments to the space of $\mathcal{E}(\mathcal{X})$. As we may assume that the players have access to the payoff functions, constructing a map between $X_t$ and the equilibrium value process can be relatively easier.
Indeed, due to the Markovian property of $X_t$, it turns out we only need a sequence of $\mathcal{B}(\mathbb{R})/\mathcal{G}$-measurable value functions $\{ v_t(\cdot) \}_{t\in [T]}$ that satisfies the following conditions:
\begin{equation}
    \begin{aligned}
          v_T (x) & = \zeta(x) , \quad \text{ for all } x \in \mathcal{X}, \\ 
        v_{t} (x) & \in \operatorname{SE}\left[\begin{array}{cc}\zeta(x) & \phi(x) \\ \psi(x) & \mathscr{T}v_{t+1} (x)]\end{array}\right] , \quad \text{ for all } x \in \mathcal{X}, t \in [T-1], 
    \end{aligned}
\end{equation}
where $\operatorname{SE}$ stands for the set of Nash (saddle-point) equilibrium values of the matrix game with two pure strategies.
Then, the last iterate value function is $\zeta(\cdot)$ by construction. The rest of the business is to figure out the backward induction equation that involves the $\operatorname{val}(\cdot)$ operator, which still relies on the calculation of $\mathscr{T}$ leveraging Monte-Carlo sampling type of methods.
Following \Cref{lemma:monotone} the monotonicity still holds, $ \{ v_t( \cdot )\}_{t\in [T]}$ is decreasing, which can be interpreted as that the decision made at the outset is most valuable, as time passes, the opportunity fades.
 For any $t \in [T]$, we define the two stopping times,
 \begin{align*}
     \tau^*_t  & = \inf \{ t \leq k \leq T | \{ v_k (X_k) = \zeta(X_k)\} \bigcup \{ v_k (X_k) = \phi(X_k)\}\} ,\\ 
    \sigma^*_t & = \inf \{ t \leq k \leq T | \{ v_k (X_k) = \zeta(X_k) \} \bigcup \{ v_k (X_k) = \psi(X_k) \}\}.
 \end{align*}

 By \Cref{thm:adcthm}, $(\tau^*_0, \tau^*_0 )$ is the equilibrium strategy pair, the definition of which reflects the consistency of value function computation, that is, the players' current value estimates either reach the early termination threshold or confrontational threshold. 
\begin{theorem}\label{thm:adcthm}
   Under ADC, the following statements hold for arbitrary initial state $x \in \mathcal{X}$: 
    \begin{itemize}
        \item for every $t \in [T]$, and all $\tau \in \T_t, \sigma \in \T_t$,
    \begin{align*}
        \mathbb{E}[ H(\tau, \sigma^*_t) |\F_t] \leq   \mathbb{E}[  H( \tau^*_t, \sigma^*_t) | \F_t]  \leq  \mathbb{E}[ H(\tau^*_t, \sigma) |\F_t] .
    \end{align*}
    \item the game admits a DDE strategy $(\tau^*_0, \sigma^*_0)$, at which the value function satisfies
    \begin{equation*}
        \begin{aligned}
            \mathcal{V}^{\tau^*_0, \sigma^*_0}(x) & =  {\essp}_{\tau \in \T} {\essf}_{\sigma \in \T}\mathcal{V}^{\tau, \sigma} (x)\\
            & = {\essf}_{\sigma \in \T} {\essp}_{\tau \in \T}  \mathcal{V}^{\tau, \sigma} (x) .
        \end{aligned} 
    \end{equation*}
    \end{itemize}
\end{theorem}

We omit the proof here as \Cref{thm:adcthm} can be seen as an extension of \Cref{thm:eqst}, to which the reasoning is similar. One can simply construct the sequence of value functions with a constant translation, and the results still hold.

\subsubsection{Case II: Defensive Dominance}
Under the Defensive Dominance Condition (DDC),  the payoff functions satisfy the ordered condition $\psi \leq \zeta \leq \phi$ for all system states $x \in \mathcal{X}$. In this case, at any state $x \in \mathcal{X}$, the defender can bide his time during the interactions of cyber kill/defense chain, as the systematic loss after the execution of Command \& Control is mitigable. Such a condition happens when the defender possesses a superior and robust position.
    This is also called \textit{second-mover advantage}, that is, the defender has the incentive to wait for the opponent to end the game.

% In this case, there is no advantage gained by stopping faster than the adversary, both parties have the intention to be engaged longer during the cyber kill/defense chain. 
DDC corresponds to the ordered payoff condition for standard Dynkin's game, where the existence and uniqueness of a saddle point value process have been proved.
The constructive sequence of (locally integrable) random variables $\{ V_t\}_{t=0}^T $, in this case, is now more straightforward (as discussed in \cite{kingman1976discrete}), defined by 
\begin{equation}
    \begin{aligned}
       & V_T   = \zeta(X_T),  \\
       & V_t   =  \min\{  \psi(X_t) , \max\{ \phi(X_t), \mathbb{E} [ V_{t+1}| \F_t]\}\} , \quad \quad \text{ for } t = 0, \ldots, T-1, 
    \end{aligned}
\end{equation}
with the stopping time strategies defined as
\begin{equation*}
    \begin{aligned}
     \bar{\tau}_t & = \inf \{  t \leq k \leq T |  V_k = \phi (X_k) \}, \\ 
     \bar{\sigma}_t & = \inf \{  t \leq k \leq T | V_k  = \psi (X_k)\} .
    \end{aligned}
\end{equation*}

  \begin{theorem}
      
  \label{thm:ch2}
    Under DDC, the following statements hold:
    \begin{itemize}
        \item[i)] for each $t \in [T]$, and for all $\tau \in \T_t$, $\sigma \in \T_t$,
        \begin{align*}
          &   V_t  = \mathbb{E}[ V_{\tau^*_t \wedge \sigma^*_t} |\F_t] = \mathbb{E}[H(\tau^*_t, \sigma^*_t)| \F_t] ,\quad\quad \text{ and, } \\ 
          &  \mathbb{E}[H(\tau, \sigma^*_t)| \F_t]  \leq   V_t \leq \mathbb{E}[H(\tau^*_t, \sigma)| \F_t]. \quad \quad  
        \end{align*}
        \item[ii)] at every time $t \in [T]$, a pair $(\tau^*_t, \sigma^*_t)$ is an equilibrium point for the subgame starting at time $t$, and the DDE value corresponding to $(\tau^*_0, \sigma^*_0)$ is 
        \begin{equation*}
             \mathbb{E}[V_0] = \mathbb{E}[V_{\tau^*_0 \wedge \sigma^*_0 }]  = \mathbb{E}[ H (\tau^*_0 , \sigma^*_0)].
        \end{equation*}
    \end{itemize}
 \end{theorem} 

  \begin{proof}
       Similar to previous results, we shall give the proof for the ``$\geq$'' side. First, we examine the trivial case where $t = T$. Obviously, there's no option but stop for both parties, so $G_t= \zeta(X_t) = \mathbb{E}[ G_{\tau \wedge \sigma^*_t} | \F_t] = \mathbb{E}[ H (\tau , \sigma^*_t) | \F_t]$ for all $\tau \in \T_T = \{ T\}$.

      Fix a $t < T$. Choose some $k$ such that $ t \leq k \leq \tau^*_t \wedge \sigma^*_t$, we have $V_k = \mathbb{E}[V_{k+1}|\F_k]$ by definition. 
      Thus, $\{ V_{ k \wedge \tau^*_t \wedge \sigma^*_t } \}_{ k = t}^T$ is a Martingale.
      Applying Doob's optional sampling theorem, one has
      \begin{equation*}
         V_t = \mathbb{E} [V_{\tau  \wedge \tau^*_t \wedge \sigma^*_t} | \F_t ], \quad \text{ for all } \tau \in \T_t. 
      \end{equation*}
      Let $\tau = \tau^*_t$, we arrive at
      \begin{align*}
        V_t & = \mathbb{E} [V_{\tau^*_t \wedge \sigma^*_t} | \F_t ] \\
         & = \mathbb{E} [V_{\tau^*_t} \mathds{1}_{\{ \tau^*_t \leq \sigma^*_t\}} + V_{\sigma^*_t} \mathds{1}_{\{ \tau^*_t >  \sigma^*_t\}} | \F_t] \\
         & = \mathbb{E} [ \phi(X_{\tau^*_t})\mathds{1}_{\{ \tau^*_t < \sigma^*_t\}} + \psi(X_{\sigma^*_t}) \mathds{1}_{\{ \tau^*_t >  \sigma^*_t\}} + \zeta(X_{\tau^*_t}) \mathds{1}_{\{ \tau^*_t =  \sigma^*_t\}}  | \F_t] 
         \\ & = \mathbb{E}[ H(\tau^*_t, \sigma^*_t)|\F_t] .
       \end{align*}
    It is also obvious that when $ t \leq k < \sigma^*_t$, then $V_k < \psi(X_k)$, therefore $V_k = \mathbb{E}[ V_{k+1} |\F_k]$.
    This implies that $ \{ V_{k \wedge \sigma^*_t} \}_{k = t}^T$ is a supermartingale. Hence, $V_t \geq \mathbb{E}[ V_{\tau \wedge \sigma^*_t} |\F_t ]$ for all $\tau \in \T_t$, 
    \begin{align*}
        \mathbb{E} [ V_{\tau \wedge \sigma^*_t} | \F_t ] & \geq \mathbb{E}[ \phi(X_{\tau})\mathds{1}_{\{ \tau < \sigma^*_t\}} + \psi(X_{\sigma^*_t}) \mathds{1}_{\{ \tau^*_t >  \sigma^*_t\}} + \zeta(X_{\tau}) \mathds{1}_{\{ \tau =  \sigma^*_t\}}  | \F_t] \\
        & = \mathbb{E}[H(\tau, \sigma^*_t) | \F_t], 
    \end{align*}
   since $V_k > \phi(X_k )$ and $\zeta(X_k) \leq \psi(X_k)$ for all $0 \leq k \leq T$. Claim ii) follows immediately.
    
   \end{proof}

Again we generalize the result to $\mathcal{E}(\mathcal{X})$, we wish to find a sequence of $\mathcal{B}(\mathbb{R})/\mathcal{G}$-measurable functions $\{ v_t(\cdot) \}_{t\in [T]}$ that satisfies the following conditions (or being shifted by a constant):
\begin{equation*}
    \begin{aligned}
       & v_T(x)   = \zeta(x) ,  \quad\quad  \text{ for all } x \in \mathcal{X} , \\
       & v_t (x)  =  \min\{  \psi(x) , \max\{ \phi(x), \mathscr{T} v_{t+1} (x )\}\} , \quad \quad \text{ for all } x \in \mathcal{X}, t = [T-1] ,
    \end{aligned}
\end{equation*}
and the DDE pair $(\tau^*, \sigma^*)$ can be defined as:
\begin{equation*}
    \begin{aligned}
        \tau^* & = \inf\{ k \in [T] | v_k(X_k ) = \phi(X_k)\} , \\ 
        \sigma^* &= \inf  \{ k \in [T] | v_k(X_k) = \psi(X_k)\} .
    \end{aligned}
\end{equation*}

\begin{theorem}
    Under ADC, the game admits a DDE strategy pair $(\tau^*, \sigma^*)$, such that
    \begin{equation*}
        \begin{aligned}
            \mathcal{V}^{\tau^*, \sigma^*}(x) & =  {\essp}_{\tau \in \T} {\essf}_{\sigma \in \T}\mathcal{V}^{\tau, \sigma} (x)\\
            & = {\essf}_{\sigma \in \T} {\essp}_{\tau \in \T}  \mathcal{V}^{\tau, \sigma} (x),
        \end{aligned}
    \end{equation*}
    for all $x \in \mathcal{X}$.
\end{theorem}

Under DDC, the optimal strategies for the players are waiting for the equilibrium process to hit the lower/upper bound of the payoff values.

\subsubsection{Decision Dominance with Information Asymmetry}

In the MDW scenarios, it is crucial to recognize that both defenders and attackers operate within an environment of information asymmetry \cite{huang2019dynamic,pawlick2019game}. This is particularly evident when considering STIX logs, as the information accessible to attackers differs from what defenders can observe. 
While defenders have the advantage of comprehensive logs that capture security events and indicators of compromise, attackers possess their own set of advantages stemming from their ability to exploit the gaps in the defender's knowledge. Attackers can leverage their insider information, external reconnaissance, and targeted intelligence gathering to gain insights into the defender's security measures, potential vulnerabilities, and defensive capabilities. In the meantime, the defender may have deceptive defense mechanisms that hide their tactics, techniques, and procedures (TTPs), to counteract the malicious exploitation.

To formalize the notion, we redefine $(X_t)_{t=0}^T$ as the true system state (which cannot be completely captured by the STIX logs), and let $(O^i_t)_{0 \leq t \leq T} (i = 1, 2)$ be the observation process for the defender ($i=1$) and the attacker ($i=2$), which jointly live in the space $(\mathcal{O}^1 \times \mathcal{O}^2, \mathcal{H}^1 \otimes \mathcal{H}^2)$, adapted to the filtrations $\mathbb{H}^1 = ( \mathcal{H}^1_t)_{0 \leq t \leq T}$ and $\mathbb{H}^2 = ( \mathcal{H}^2_t)_{0 \leq t \leq T}$.
This information asymmetry enables the players to make informed decisions regarding their strategies, tactics, and the selection of attack vectors/defensive mechanisms. 
Therefore, defenders must not only rely on STIX logs and robust defense mechanisms but also proactively bridge the information gap by enhancing their threat intelligence capabilities, anticipating adversary behaviors, and continuously evolving their defense strategies to counter the advantages of information asymmetry in the cyber landscape.

To formally define the DDG under asymmetric information structure, we denote by $\T(\mathbb{H}^i)$ the set of $\mathbb{H}^i$-stopping times, $\T(\mathbb{H}^i) = \{0 \leq \tau \leq T:  \{\tau(\omega) \leq k\} \in \mathcal{H}^i_k  \ \forall k \in [T], \forall \omega \in \Omega  \}$.
The decision payoffs at each stage $t \in [T]$, in this case, may depend on both $O^i_t$ and $X_t$. 
% Specifically, we assume that $\phi, \zeta, \psi \in \mathcal{E}(\mathcal{X} \times \mathcal{Y})$, with a dependency on either $X_t$ or $Y_t$, or both of them.
Following the standard formalism of the Partially Observable Markov Decision Process (POMDP), we assume that the payoff functions still only depend on the true system state, which is a hidden latent variable for both players. Instead, there exists an emission kernel $\mathbb{O}: \mathcal{X} \to \Delta(\mathcal{O}^1 \times \mathcal{O}^2)$ that measures the joint probability of observations made by the defender and the attacker.
An illustration is shown in \Cref{fig:ddgwai}

\begin{figure}
    \centering
    \includegraphics[width =\textwidth]{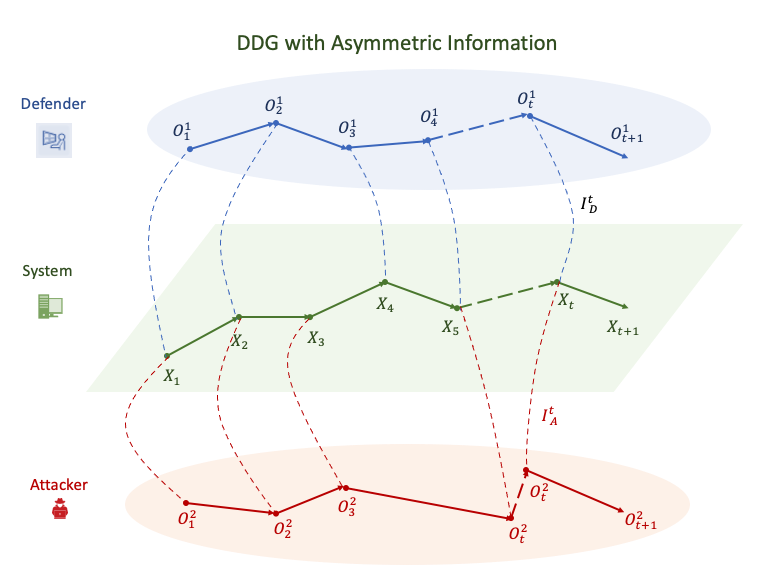}
    \caption{An illustration of asymmetric information dynamic games defined in \Cref{def:ddgasym}. The two players have distinct partial
observations for the system state $X_t$, denoted by $(O^1_t, O^2_t)$. In DDG,
the defender has to infer the true state to determine the stopping time strategy based on 
the payoff structure, which relies on credible modeling, requiring expertise in the fundamental understanding of the cyber threats}
    \label{fig:ddgwai}
\end{figure}
% For instance, when system state variables are only partially observable by the defender, who has access to her own early termination payoff function $\phi(\cdot): \mathcal{X} \to \mathbb{R}$, and confrontation payoff function $\zeta(\cdot, \cdot): \mathcal{X} \times \mathcal{Y} \to \mathbb{R}$. The attacker, on the other hand, only has access to and confrontation payoff and her own termination payoff function $\psi(\cdot): \mathcal{Y} \to \mathbb{R}$.

Factorization Lemma says in order to infer the true states from the partial observations, say, if $O^1_t$ is $\F_t$/$\mathcal{H}^1_t$-measurable, there needs to be a deterministic $\F_t$/$\mathcal{H}^1_t$-measurable map $f: \mathcal{X} \to \mathcal{O}^1$ such that $O^1_t = f(X_t)$, whose existence and accessibility are not always guaranteed in the cyber domain. 
Therefore, it is reasonable to assume that the players have their stopping time strategies restricted to $\T(\mathbb{H}^i)$. \Cref{def:ddgasym} summarizes the game under asymmetrical information structure.

\begin{svgraybox}
     \begin{definition}[Decision Dominance Game with Information Asymmetry] \label{def:ddgasym}
          A tuple $( \mathcal{X}, \mathcal{O}^1 \times\mathcal{O}^2, \Pk, \mathbb{O}, \phi, \zeta, \psi, \mathcal{T}(\mathbb{H}^1), \mathcal{T}(\mathbb{H}^2))$ encapsulates a Decision Dominance Game with Information Asymmetry (DDGIA) if it satisfies that
          \begin{itemize} 
              \item There exists a hidden Markov process $(X_t)_{0 \leq t \leq T}$ that lives in $(\mathcal{X}, \mathcal{G})$ with transition kernel $\Pk$, which yields observations $(O^1_t, O^2_t)$ through emission kernel $\mathbb{O}$;
              \item $\phi, \zeta,$ and $\psi$ are payoff functions mapping from $X_t$ to $\mathbb{R}$, $ \phi, \zeta, \psi  \in \mathcal{E}(\mathcal{X})$, which is the set of all bounded $\mathcal{B}(\mathbb{R})/\mathcal{G}$-measurable functions on $(\mathcal{X}, \mathcal{G})$. Further, $\min(\phi, \psi) \leq \zeta \leq \max(\phi, \psi)$ on $\mathcal{X}$;
              \item At each stage $t$, player $i$ ($i=1, 2$) picks a stopping strategy from space $\T_t(\mathbb{H}^i):= \{ t \leq \tau \leq T:  \{\tau(\omega) \leq k\} \in \mathcal{H}^i_k  \  \forall k \in [T], \forall \omega \in \Omega  \}$ to decide whether to stop or continue the kill/defense chain.
              \item At each stage the utility function of the defender is 
              \begin{equation*}
                   H(\tau_t, \sigma_t) = \phi(X_{\tau_t}) \mathds{1}_{\{ \tau_t < \sigma_t\}} + \zeta(X_{\tau_t}) \mathds{1}_{\{ \tau_t = \sigma_t\}} +  \psi(X_{ \sigma_t })\mathds{1}_{\{ \tau_t >  \sigma_t\}}, 
              \end{equation*}
              while the attacker attains $-H(\tau_t, \sigma_t)$.
          \end{itemize}
     \end{definition}
\end{svgraybox}

The goal of the defender is to choose $\tau$ to maximize her utility under all possible choices of the attacker, which leads to the lower value function of DDGIA, 
\begin{equation}
     \underline{V}(x) = \essp_{\tau \in \T(\mathbb{H}^1)} \essf_{\sigma \in \T(\mathbb{H}^2)} V^{\tau, \sigma} (x) .
\end{equation}
Similarly, the goal of the attacker is to choose $\sigma$ to minimize the defender's utility under all possible choices of the defender, which leads to the upper-value function, 
\begin{equation}
     \overline{V}(x) =  \essf_{\sigma \in \T(\mathbb{H}^2)} \essp_{\tau \in \T(\mathbb{H}^1)} V^{\tau, \sigma} (x) .
\end{equation}
\begin{svgraybox}
    \begin{definition}[DDE with Information Asymmetry] \label{def:neia}
    A pair of stopping time strategies $(\tau^*, \sigma^*) \in \mathcal{T}(\mathbb{H}^1) \times \mathcal{T}(\mathbb{H}^2) $ is a Decision-Dominance Equilibrium (DDE) if for all initial state $x \in \mathcal{X}$, it satisfies the minimax condition:
    \begin{equation} 
    \label{eq:ddeia}
\begin{aligned}
    V^{\tau^*, \sigma^*} (x) & = \essp_{\sigma \in \T(\mathbb{H}^2)}  \essf_{\tau \in \T(\mathbb{H}^1)} V^{\tau, \sigma}(x) \\ 
       & = \essf_{\tau \in \T(\mathbb{H}^2)} \essp_{\sigma \in \T(\mathbb{H}^2)}  V^{\tau, \sigma}(x).
\end{aligned}
\end{equation} 
\end{definition}
\end{svgraybox}

We say that a DDGIA has a value if $\underline{V}(x) = \overline{V}(x)$. 
Note that the existence and uniqueness of the value is a non-trivial question in general, as we shall find the reasoning presented in the previous section not applicable due to the introduction of two private filtrations for both parties. 
In principle, the value exists if $\mathbb{H}^i$ reveal the same information from $\mathbb{F}$, in which case the conditional expectation $\mathbb{E}(\cdot | \mathbb{H}^i_t)$ can be seen equivalent with $\mathbb{E}(\cdot | \F_t)$, thus the players will make their decisions using the same threshold policies. 
This property, however, requires some special structures of the observation kernel $\mathbb{O}$, which might not hold in realistic scenarios.

\subsection{Decision Dominance Zero-Trust Defense (DD-ZTD): A Case Study} 

In this case study, we consider an $T$-episodic DDG with symmetric information over the same 5G network $G= \langle V, E\rangle$ as discussed in section \ref{subsec:lateral}, where each episode $t$ contains $H$ ZTD steps against lateral movement. 
The ZTD state action variables within one episode $t$ is $\mathbf{sa}_t = (s^1_t, a^1_t, s^2_t, a^2_t \ldots, a^{H-1}_t, s^H_t )$, where $s^h_t = (G^h_t, L^h_t), h = 1, \ldots, H$ are the authentication graphs and the visiting indicator functions at episode $t$, and the joint actions $a^h_t$ are automated by the threshold-policy trust engine, which is either the Bayesian type or the Machine Learning type. 
Denote the STIX logs within $t$ as $x_t \in \mathcal{X}$, which includes but is not limited to the events of 5G network exposure, slicing control, session management; the threat actor characterizations such as suspected user intentions and handling guidance. 
The Markovian state at episode $t$ is a composition of both historical ZTD state action variables and the STIX logs gathered before episode $t$, i.e., $X_t: = (\mathbf{sa}_{1:t-1}, x_{t-1})$. 

During the cyber kill/defense chain interaction, at the beginning of each episode, the defender can choose to completely cut off the chain before episode $t$ starts by isolating the networks, restarting the services, resetting all the credentials, patching and hardening the security configurations, and then restoring and resuming the operations. The cost of the defender's cutting-off strategy is $C(\cdot): \mathcal{X} \to \mathbb{R}$, which only depends on the cyber threat information.
Similarly, the attacker can choose to take action early by exploiting Zero-Day vulnerabilities, evading intrusion detection systems, and implementing stealthy command and control at an early stage of the cyber kill chain. Again we let the exploitation loss be $\ell(\cdot): \mathcal{X} \to \mathbb{R}$, which completely depends on the cyber threat characterization of episode $t$. Now we are ready to define the three payoff functions in our DDG framework. 

The early termination payoff, confrontation payoff, and late termination payoff functions can be defined as
\begin{equation}
\begin{aligned}
     \phi(X_t) & = - \mathbb{E}[\sum_{k=1}^{t-1} \sum_{h =1}^H u_D(s^h_k, a^h_k)] - C(x_{t-1}), \\
    \zeta(X_t) & =  - \mathbb{E}[\sum_{k=1}^{t-1} \sum_{h =1}^H u_D(s^h_k, a^h_k)] - C(x_{t-1}) - \ell(x_{t-1}), \\
     \psi(X_t) & = - \mathbb{E}[\sum_{k=1}^{t} \sum_{h =1}^H u_D(s^h_k, a^h_k)] - \ell(x_{t-1}),
\end{aligned}
\end{equation}
where the expectation $\mathbb{E} [\sum_{h}^H u_D(s^h_t, a^h_t)]$ is taken conditioned on $\mathbf{sa}_{1:t-1}$.
The interpretation is that when the defender chooses to shut down and restore the services, the ZTD stops for that episode, while if the attacker chooses to exploit early, the ZTD mechanism is still active. 

One can easily verify that when both $\ell$ and $C$ are positive and the expected ZTD cost within every episode $t$ satisfies $\mathbb{E} [\sum_{h}^H u_D(s^h_t, a^h_t)] > C(x_{t-1})$ the DDG satisfies DDC.

}

% {
% \section{Case Study}
% \label{sec:case}
% \input{case.tex}
% }

{
\section{Conclusion}
This chapter develops a game-theoretic framework for the decision-dominant zero-trust defense of 5G networks in the face of advanced persistent threats that utilize a cyber kill chain to disrupt the network operation. The advanced features of 5G networks, despite their contributions to multi-domain integration, bring a larger attack surface and render the network system vulnerable in the presence of advanced persistent threats (APT) and other malicious attacks. The combination of vulnerabilities in APT, supply chains of 5G equipment, and network slicing, along with others, can be exploited by an APT attacker to create a cyber kill chain consisting of reconnaissance, planning, execution, and exploration.  

To outmaneuver the malicious attacker and thwart the kill chain, this chapter proposes a decision-dominant zero-trust defense (DD-ZTD) framework, a proactive defense mechanism enabling the defender to make timely and effective decisions with incomplete information regarding the situation and disrupt the kill chain before its completion. Two pillars of DD-ZTD are game-theoretic zero-trust defense built upon asymmetric information Markov games (AIMG) and decision-dominance defense characterized by Dykin's stopping-time games. With the AIMG's expressive power on information structures in cyber defense, ZTD employs a variety of trust engines to evaluate entities' trustworthiness with limited partial observations, which is then fed into the access policy powered by equilibrium thinking that anticipates the attacker's response. We further present an end-to-end ZTD facilitated by recent machine learning advancements with data-driven trust evaluation and explainable and generalizable policy learning. 

While the proposed ZTD offers a set of fruitful tools to quantitatively analyze trustworthiness under information asymmetry, the networked entities still face multi-stage persistent cyber threats that call for rapid response from the defender. To outpace the attacker's kill chain, decision-dominance defense (D$^3$), mathematically treating interactions of cyber defense/kill chain as a stopping-time game, aims to take the decisive move to cut off the kill chain before the attack materializes. The essence of D$^3$ is the timing of the cutting-off, which is determined by the equilibrium of the game with anticipation of the attacker's strategic move. The resulting DD-ZTD, as an organic integration of the two game-theoretic defense mechanisms, displays great potential in combating sophisticated adversaries, which we articulate using a case study in 5G network defense.          
}

%%%%%%%%%%%%%%%%%%%%%%%% referenc.tex %%%%%%%%%%%%%%%%%%%%%%%%%%%%%%
% sample references
% %
% Use this file as a template for your own input.
%
%%%%%%%%%%%%%%%%%%%%%%%% Springer-Verlag %%%%%%%%%%%%%%%%%%%%%%%%%%
%
% BibTeX users please use
\bibliographystyle{spbasic-unsort}
\bibliography{ref.bib}

\end{document}